\documentclass[aps,prx,groupedaddress,twocolumn,showpacs,longbibliography,10pt]{revtex4-2}

\usepackage[table]{xcolor}
\usepackage{amsmath}
\usepackage{slashed}
\usepackage{amssymb}
\usepackage{footmisc}

\usepackage{graphicx}
\usepackage{draft}
\usepackage{color}
\usepackage{times, verbatim}      
\usepackage{mathtools,tikz}

\usetikzlibrary{arrows.meta,calc,decorations.pathmorphing,positioning}

\usepackage{pgfplots}
\usetikzlibrary{decorations.pathmorphing}
\usetikzlibrary{matrix}
\usepackage{natbib}
\usepackage{multirow}
\usepackage{hyperref}

\definecolor{colour1}{rgb}{0.368417, 0.506779, 0.709798}
\definecolor{colour2}{rgb}{0.880722, 0.611041, 0.142051}
\definecolor{colour3}{rgb}{0,1,1}
\definecolor{colour4}{rgb}{0,1,0}
\definecolor{colour5}{rgb}{1,1,0}
\usetikzlibrary{decorations.pathreplacing}
\usepackage{dsfont}

\usepackage{amsthm}

\hypersetup{
       colorlinks=true,
}

\usepackage{multirow}
\makeatletter
\newsavebox{\@brx}
\newcommand{\llangle}[1][]{\savebox{\@brx}{\(\m@th{#1\langle}\)}%
  \mathopen{\copy\@brx\kern-0.5\wd\@brx\usebox{\@brx}}}
\newcommand{\rrangle}[1][]{\savebox{\@brx}{\(\m@th{#1\rangle}\)}%
  \mathclose{\copy\@brx\kern-0.5\wd\@brx\usebox{\@brx}}}
\makeatother

\newcommand{\ep}{\epsilon}

\newcommand{\pa}{\partial}
\newcommand{\CO}{{\cal O}}
\newcommand{\CN}{{\cal N}}
\newcommand{\CI}{{\cal I}}
\newcommand{\SO}{{\text{SO}}}

\begin{document}
 
\author{Oleksandr Diatlyk$^{1}$, Zimo Sun$^{2,3}$, and Yifan Wang$^{1}$
}
\affiliation{$^{1}$ Center for Cosmology and Particle Physics, New York University, New York, NY 10003, USA}
\affiliation{$^{2}$ Institute for Advanced Study, Princeton, NJ 08540, USA}
\affiliation{$^{3}$ Joseph Henry Laboratories, Princeton University, Princeton, NJ 08544, USA}

\title{ \hfill{\normalfont\small PUPT-2661}\\[2em]
Extraordinary Surface Criticalities for Interacting Fermions
}

\begin{abstract}

Interacting fermions exhibit a rich landscape of surface defects and associated critical phenomena. We investigate novel surface critical behavior in the three-dimensional Gross–Neveu–Yukawa model. For a class of defect renormalization group flows, we obtain exact infrared solutions and show how fermionic anomalies are encoded in the resulting surface dynamics. We further uncover emergent topological and geometric structures in the defect coupling space, and comment on their relation to a defect analogue of the CFT distance conjecture.

\end{abstract}

\date{\today}


\maketitle

\section{Introduction and Summary}
Strongly coupled systems exhibit novel critical behavior in the presence of surface impurities, driven by nontrivial bulk–defect interactions and often characterized by large anomalous dimensions and emergent defect degrees of freedom. A canonical example is the $3d$ ${\rm O}(N)$ Wilson–Fisher model, whose boundary and interface deformations give rise to a rich phase structure beyond the bulk critical point. By tuning the ${\rm O}(N)$-invariant surface coupling, the model realizes the ordinary, special, extraordinary, and extraordinary-log universality classes, each with distinct boundary critical behavior \cite{Bray:1977fvl,McAvity:1995zd,Cardy_1996,Diehl:1996kd,Metlitski:2020cqy,Giombi:2020rmc,Padayasi:2021sik,Toldin:2021kun,Krishnan:2023cff}.
In particular, 
surface defects \cite{Krishnan:2023cff,Trepanier:2023tvb,Giombi:2023dqs,Raviv-Moshe:2023yvq,Diatlyk:2024ngd} generated by the leading singlet operator on the trivial surface admit an infrared (IR) factorization into boundary conditions dressed by gapless surface modes. This structure is fixed by symmetry and is consistent with defect renormalization group (RG) monotones. It was shown in \cite{Popov:2025cha} that such defects, termed ``generalized pinning fields,’’ universally obey this factorization, providing a unified description of these boundary universality classes.

Fermionic systems, ubiquitous in both high-energy and condensed-matter physics, exhibit an even richer structure due to the wide array of anomalies they can carry \cite{Kapustin:2014dxa,Witten:2015aba,Seiberg:2016rsg,Freed:2016rqq,Witten:2016cio}. 
These anomalies constrain the infrared dynamics and often enforce gapless modes on edges and domain walls \cite{Jackiw:1975fn,Callan:1984sa}. Fermionic degrees of freedom also enable new classes of defects. A natural question is therefore how bulk anomalies are encoded on such defects and how they constrain defect dynamics. For surface defects in $3d$, this question is particularly sharp, since their extended worldvolume can enclose both local and extended operators and is thus especially sensitive to anomaly data. Moreover, $3d$ fermionic systems are believed to form a rich web of dualities \cite{Seiberg:2016gmd,Metlitski:2016dht}, and defects provide a dispensable probe for testing and refining these dualities which are starting to be explored \cite{Komargodski:2025jbu}. 

Determining the infrared dynamics of defects requires analyzing defect renormalization group flows, which is notoriously challenging. In particular, few results are available for defects in interacting fermionic theories in $3d$ without supersymmetry (see \cite{Giombi:2021cnr,Herzog:2022jlx,Giombi:2022vnz,Jiang:2025sfb,Fedorenko:2026svv} for recent progress). In this work, we obtain exact conformal solutions for strongly coupled surface defects in a prototypical interacting fermionic theory, the $3d$ Gross–Neveu–Yukawa (GNY) model, which describes parity-breaking transitions with dynamical fermion mass generation (e.g., semimetal–insulator transitions in graphene). The surface defect we introduce is of the generalized pinning type \cite{Popov:2025cha}, defined by a surface perturbation sourced by the normal derivative of the leading pseudoscalar operator. As we explain, the general factorization result of \cite{Popov:2025cha} plays a central role in identifying the infrared solutions.

The GNY model consists of $N$ Majorana fermions $\Psi^I$ coupled to a real pseudoscalar $\phi$ via the following Lagrangian,
\ie
\label{GNYL}
L_{\rm GNY}=-\frac{1}{2}(\partial \phi)^2 -\frac{1}{2}\bar{\Psi}^I (\slashed{\partial}+g_1\phi) \Psi^{I}- \frac{\kappa}{2} \phi^2 -\frac{g_{2}}{4!} \phi^4 \,,
\fe
and has a global ${\rm O}(N)$ symmetry whose center coincides with fermion parity $(-1)^F$. It also admits parity and time-reversal symmetries with intricate anomalies \cite{Witten:2015aba,Seiberg:2016rsg,Witten:2016cio}. The theory flows to an interacting fixed point, the GNY CFT, where these symmetries are preserved by the vacuum. This CFT has been extensively studied using the conformal bootstrap \cite{Iliesiu:2015qra,Erramilli:2022kgp, Mitchell:2024hix}, $\epsilon$-expansion \cite{Fei:2016sgs, Zerf:2017zqi, Gracey:2025}, and large-$N$ methods \cite{Gracey:1990wi,Gracey:1992cp, Vasiliev:1993,Gracey:1993kc}, with all evidence indicating that the pseudoscalar satisfies $\Delta_\phi<1$ for all $N$.

Motivated by this, we consider the following ${\rm O}(N)$-invariant surface deformations to \eqref{GNYL} supported at $z=0$ (with worldvolume coordinates $(t,x)$),
\ie
L_{\rm defect}= (h_{\rm o} \phi + h_{\rm m} \partial_z \phi + h \phi^2)\delta(z)\,,
\label{LDefect}
\fe
which define a three-parameter family of generalized pinning defects \cite{Popov:2025cha} in the GNY CFT. This space of defects exhibits a rich structure of fixed points and phases, depending nontrivially on $N$. In this work, we focus on the case with bare couplings set to $h_{\rm o}=h=0$, which we refer to as the\textit{ extraordinary surface defect}, which sharply exposes the interplay between bulk anomalies and surface dynamics. Additional structures across the full defect phase diagram \eqref{LDefect} will be explored in companion works \cite{SZY26, Ord}. In Appendix~\ref{app:norm}, we also solve the case with bare couplings $h=h_{\rm m}=0$, which defines the \textit{normal surface defect} and comment on its relation to the extraordinary surface via a nontrivial defect fusion algebra.

A main result of this paper is an exact IR description of the extraordinary surface defect. Depending on the sign of $h_{\rm m}$, the defect factorizes into a conformal boundary condition of the GNY CFT—referred to as the \textit{normal boundary condition} in analogy with the Ising model \cite{Diehl:1996kd}—together with \textit{emergent decoupled} $2d$ chiral or anti-chiral Majorana fermions denoted by  $\chi^I$ and $\tilde\chi^I$ respectively. Concretely, in Euclidean signature, 
\ie
\begin{array}{@{}r@{\hspace{0.8em}}c@{\hspace{1em}}l@{}}
\multirow[c]{2}{*}[0ex]{$\cD_{\rm ext} \equiv \bigl[e^{h_{\rm m}\int_{z=0}\partial_z\phi}\bigr]_{\rm ren}$}
& \raisebox{0ex}{\smash{\rotatebox[origin=c]{10}{$\xrightarrow{\;h_{\rm m}>0\;}$}}}
& |B_1\rangle\langle B_1|+\chi^I \\[1ex]
& \raisebox{0ex}{\smash{\rotatebox[origin=c]{-10}{$\xrightarrow{\;h_{\rm m}<0\;}$}}}
& |\overline{B_1}\rangle\langle \overline{B_1}|+\tilde\chi^I
\end{array}
\label{mainsol}
\fe
Here $|B_1\rangle$ denotes the normal boundary condition, which is mapped to $\langle B_1|$ by transverse parity and to $|\overline{B_1}\rangle$ by longitudinal parity (or time reversal). The flows preserve transverse parity, and the two branches are related by time reversal.  

As we explain in Section~\ref{sec:anomaly}, this result follows from combining the nonperturbative factorization property of generalized pinning defects \cite{Popov:2025cha}, prior knowledge of conformal boundary conditions of the GNY CFT \cite{Giombi:2021cnr}, and constraints from symmetries and anomalies \cite{Witten:2015aba,Seiberg:2016rsg,Witten:2016cio}. In particular, the symmetry properties and anomalies of the defect are inherited from those of the bulk together with the specific pinning deformation in \eqref{mainsol}. The emergent decoupled (anti)chiral fermions play a crucial role in anomaly matching, and can be understood intuitively by ``squeezing'' a domain-wall profile for the fermion mass in a strip (see Fig.~\ref{fig:squeeze}), thereby generating a localized flux for the mass coupling, or heuristically a monodromy defect for the corresponding $(-1)$-form symmetry \footnote{Usual monodromy defects are defined for $0$-form symmetries, but their extension to $p$-form symmetries is straightforward: they correspond to localized flux insertions for the associated background gauge fields. For example, a non-genuine ’t Hooft loop in $4d$ ${\rm SU}(N)$ gauge theory can be viewed as a monodromy defect for the $\mZ_N$ electric one-form symmetry.
The novelty in the $p=-1$ case is that the monodromy is intrinsically tied to the gravitational anomaly localized on the defect (interface) by the anomaly (mixed with gravity) in the coupling space. In particular, it can be tuned by adding (anti)chiral modes on the defect worldvolume, which shift the anomaly and thereby modify the effective monodromy (see Section~\ref{sec:anomaly} for details).}. We further find consistency with the defect RG monotone: the $b$-function, which measures the Weyl anomaly of the conformal surface, decreases monotonically along the flow \cite{Jensen:2015swa, Casini:2018nym, Wang:2020xkc, Shachar:2022fqk}.

In Section~\ref{sec:largeN}, we solve the extraordinary surface defect explicitly at large $N$ and confirm the proposal \eqref{mainsol}. To leading order, we find two branches of continuous solutions that join into an \textit{emergent conformal manifold} with the topology of a figure eight in defect coupling space (see Fig.~\ref{fig:eight}), preserving transverse parity and ${\rm O}(N)$ symmetry throughout. The $2d$ (anti)chiral fermions appear at the tips of the figure eight and are reabsorbed through interactions with the normal boundary conditions, as we demonstrate using conformal perturbation theory in Section~\ref{sec:defspace}. We further comment on properties of this large-$N$ conformal manifold, such as the Zamolodchikov distance, its relation to the defect operator spectrum and degeneration properties at the tips, in light of a defect version of the CFT distance conjecture \cite{Ooguri:2006in,Ooguri:2024ofs}. Upon including ${1/ N}$ corrections, this conformal manifold is lifted except for the tips (confirming \eqref{mainsol}) and the crossing at the origin (trivial surface). This analysis makes explicit the physical picture that the extraordinary surface defect acts as a pump for $2d$ (anti)chiral fermions driven by local perturbations in a non-chiral gapless bulk.

In a companion paper \cite{SZY26}, we study extraordinary surface defects in the $N=1$ GNY CFT, which is known to exhibit emergent $\cN=1$ superconformal symmetry \cite{Grover:2013rc,Fei:2016sgs,Rong:2018okz,Atanasov:2018kqw,
Atanasov:2022bpi}, and further confirm \eqref{mainsol}. There, we find that the extraordinary surface defect is in fact half-BPS, preserving $\cN{=}(0,1)$ or $\cN{=}(1,0)$ supersymmetry. Interestingly, this surface supersymmetry is spontaneously broken in the IR, and the emergent (anti)chiral fermion is identified as the goldstino. Moreover, the SUSY breaking can be diagnosed via a generalization of deformation invariants for $\cN=(0,1)$ supersymmetric QFTs in terms of topological modular forms and their modules \cite{Gukov:2025nmk}.

\section{Factorization, Anomalies and Emergent Surface Chiral Fermions}
\label{sec:anomaly}

The IR factorization property established for generalized pinning defects \cite{Popov:2025cha}, when applied to the extraordinary surface defect in \eqref{mainsol}, implies that its infrared limit is described by conformal boundary conditions of the GNY CFT, possibly dressed by emergent gapless modes localized on the surface. The goal is to use symmetry and anomaly constraints to determine the precise factorization channels.

Recall the symmetry group of the GNY model
\ie 
G_{\rm sym}={\rm Pin}_+(1,2)\times_{\mZ_2^F} {\rm O}(N)
\label{GNYsym}\,,
\fe
contains time-reversal $T$ and transverse parity $P$ which act as
\ie 
 \Psi(t,x,z)\xrightarrow[]{T}  \gamma^0\Psi(-t,x,z)\,,~\Psi(t,x,z) \xrightarrow[]{P} \gamma^1\Psi(t,x,-z)\,, 
\label{parity}
\fe
on the fermions and accordingly on the pseudoscalar $\phi$. Here $T^2=(-1)^F$ is identified with the center of the ${\rm O}(N)$ symmetry. The extraordinary surface breaks $T$ explicitly since $\phi$ is parity-odd but preserves the transverse parity $P$.
 
By a surface version of the Coleman's theorem \cite{Cuomo:2023qvp}, continuous global symmetry cannot be spontaneously broken and thus at this stage, we already know that the IR extraordinary surface is described by the factorized interface $|B\ra\la B|$ for an ${\rm O}(N)$-invariant boundary condition $|B\ra$ and its partner $\la B|$ under reflection $P$, up to gapless (and topological) surface degrees of freedom \footnote{The extraordinary-log scenario in the ${\rm O}(N)$ Wilson-Fisher theory \cite{Krishnan:2023cff,Cuomo:2023qvp} does not appear to happen here as in confirmed by our direct large $N$ analysis in Section~\ref{sec:largeN}.}.

The ${\rm O}(N)$-invariant conformal boundary conditions were studied in \cite{Giombi:2021cnr} using $\ep$-expansion and large $N$. Three types of boundary conditions were found, corresponding to special $|B_2'\ra$, ordinary $|B_2\ra$ and normal $|B_1\ra$, ranked from the least stable to the most stable types \footnote{In fact, the phase diagram of ${\rm O}(N)$-invariant boundaries in the GNY model as presented in \cite{Giombi:2021cnr} is incomplete in $d=3$. In particular, there is in addition, another extraordinary boundary universality class with decoupled fermions, much like the extraordinary surface defect considered here; furthermore, the special and ordinary boundaries disappear at a critical value of $N$. Since these subtleties do not affect the most stable normal boundary condition, we will not elaborate on them in the main text and defer the detailed analysis to \cite{SZY26}.}. Only the normal boundary $|B_1\ra$ is consistent with the defect RG monotone \cite{Giombi:2021cnr}: since the UV $b$-function vanishes for the flow in \eqref{mainsol}, the putative conformal boundary $|B\ra$ must satisfy $b_{B}<0$. The normal boundary is defined by a Nahm-type singularity for the pseudoscalar and a standard projection on the $3d$ fermions,
\ie 
|B_1\ra:\phi \xrightarrow[]{z\to 0^+} {c/ z^{\Delta_\phi}}\,,~\gamma^1\Psi |_{z=0}=-\Psi |_{z=0}\,,
\label{B1bc}
\fe
where $c>0$ is a constant that is fixed by the dynamics. It describes the attractor for boundary RG flows triggered by $\phi$ \cite{Giombi:2021cnr}.
A direct large $N$ analysis in Appendix~\ref{app2} confirms that
\ie 
b_{B_1}=-{N\over 8} -{9\over 16} + O\left({1\over N} \right)\,.
\label{bB1}
\fe
The same conclusion on negative $b$ holds with existing results from $\ep$-expansion \cite{Giombi:2021cnr, DGS}. 

Having used the ${\rm O}(N)$ symmetry to limit the outcomes of the defect RG, we have almost reached the RHS of \eqref{mainsol}. What remains is to understand potential gapless surface modes that may emerge under the RG. Here is where the more subtle discrete symmetries in \eqref{GNYsym} play an important role, amplified by their nontrivial anomalies. In fact, the (global) time-reversal anomaly \cite{Witten:2016cio} already places stringent non-perturbative constraint on the ${\rm O}(N)$-invariant boundaries by requiring $T$ to be broken on the boundary \cite{Thorngren:2020yht}, which is confirmed by the explicit constructions in \cite{Giombi:2021cnr} including $|B_1\ra$ in \eqref{B1bc}. 

Furthermore, Majorana fermions in $3d$ have a mod 2 parity anomaly \cite{Redlich:1983dv,Niemi:1983rq,Alvarez-Gaume:1984zst} that obstructs its definition on orientable manifolds consistent with diffeomorphism invariance and locality while preserving parity (equivalently time-reversal) \cite{Witten:2015aba}. This anomaly can be cancelled via the inflow mechanism by coupling to a bulk topological superconductor on the four-manifold $W$ which supplies the $T$-invariant anomaly theory \cite{Seiberg:2016rsg},
\ie 
\exp\left({\pi i \over 2}\int_W \hat A(R) \right)\equiv \exp\left({i\over 384\pi} \int_W \tr R\wedge R \right)\,,
\label{parityanomaly}
\fe
also known as a gravitational theta angle $\theta_g=\pi$. Note that the $3d$ gravitational Chern-Simons term which is properly quantized on spin manifolds satisfies
\ie 
 d{\rm CS}_g ={1\over 192\pi}\tr R\wedge R\,,
\fe
and  \eqref{parityanomaly} cannot be reduced to a local counterterm, thus producing a mod 2 anomaly. 

This parity anomaly can be in fact promoted to an anomaly in the coupling space parametrized by the fermion mass $m$ which is parity-odd \cite{Cordova:2019jnf,Hason:2020yqf}. The corresponding anomaly theory is
\ie 
\exp\left({i\over 192\pi} \int_W \rho(m)\,\tr R\wedge R \right)\,,
\label{manomaly}
\fe
where $\rho(m)$ can be thought of as a $-1$-form symmetry background that satisfies $\rho(-\infty)=0$ and $\rho(\infty)=1$. The parity anomaly \eqref{parityanomaly} is recovered at $m=0$ upon imposing $T$-invariance on \eqref{manomaly} \cite{Cordova:2019jnf}. 
Moreover, given a domain wall profile of thickness $\ell$  in a spatial direction $z$,
\ie 
m(z)=m_* \tanh(z/\ell)\,,
\label{dw}
\fe 
\eqref{manomaly} produces the following $3d$ anomaly theory on $Y$,
\ie 
\exp \left(i\int_Y  {\rm CS}_g\right)\,,
\label{surfaceanomaly}
\fe
attached to the surface $\pa Y$ at $z=0$, which implies a nontrivial gravitational anomaly (chiral central charge) $c_L-c_R=1/2$ at this surface. In the free fermion theory with the mass profile $m(z)$, this can be seen easily from the localized domain wall fermion which saturates this gravitational anomaly \cite{Callan:1984sa}. However, a key point is that, the anomalies described by \eqref{manomaly} and \eqref{surfaceanomaly} are robust against interactions and thus apply to the strongly coupled GNY CFT of interest here, which is connected to the theory of $N$ free Majorana fermions by adjusting the couplings in \eqref{GNYL}. The corresponding anomaly theories (with ${\rm O}(N)$ symmetry) are given as in \eqref{parityanomaly}, \eqref{manomaly} and \eqref{surfaceanomaly} raised to the $N$-th power. 

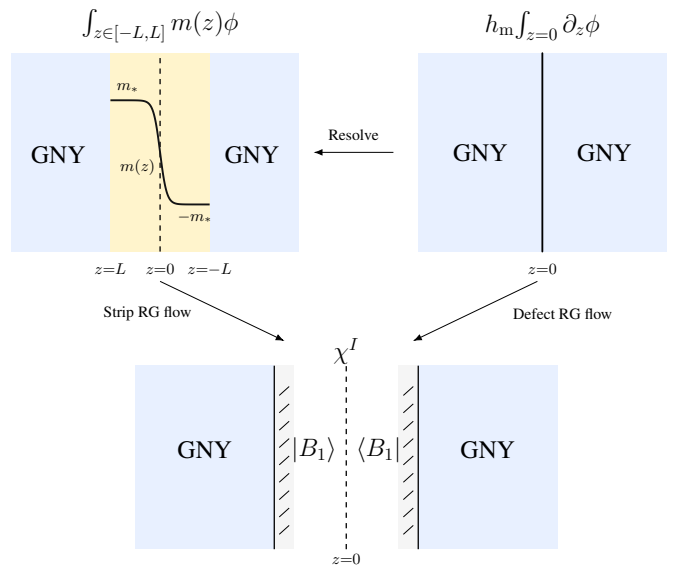
\begin{figure}[]
    \centering

 \scalebox{0.66}
 {
 \begin{tikzpicture}[
    x=1cm,y=1cm,
    >=Latex,
    every node/.style={font=\small},
    panel/.style={line width=1.05pt, line cap=round, line join=round},
    guide/.style={line width=0.75pt},
    flow/.style={-{Latex[length=3mm,width=2mm]},
    line width=0.95pt},
    hatch/.style={line width=0.45pt}
]

\definecolor{gnyfill}{RGB}{232,240,255}
\definecolor{stripfill}{RGB}{255,244,204}
\definecolor{shadefill}{RGB}{245,245,245}
\definecolor{massline}{RGB}{20,20,20}

\begin{scope}
    \fill[gnyfill] (0,0) rectangle (5.8,4.0);
    \fill[stripfill] (2.0,0) rectangle (4.0,4.0);
 
    \draw[guide,dashed] (3.0,0)--(3.0,4.0); 

    \draw[line width=1.1pt, color=massline, samples=200, domain=2.:4]
        plot (\x,{2.0 - 1.05*tanh(7.5*(\x-3.0))});
 
    \node at (0.95,1.95) {\Large GNY};
    \node at (4.85,1.95) {\Large GNY};

    \node at (3,4.55)
        {\Large $\int_{z\in[-L,L]} m(z)\phi$};

    \node at (2.55,1.7) {$m(z)$};

    \node at (3.7,0.7) {$-m_*$};
    \node at (2.35,3.3) {$m_*$};

    \node at (2,-0.35) {$z{=}L$};
    \node at (3,-0.35) {$z{=}0$};
    \node at (4,-0.35) {$z{=}{-}L$};
\end{scope}

\begin{scope}[shift={(8.2,0)}]
    \fill[gnyfill] (0,0) rectangle (5,4);
    \draw[panel] (2.5,0)--(2.5,4);

    \node at (1.25,2) {\Large GNY};
    \node at (3.75,2) {\Large GNY};

    \node at (2.5,-0.35) {$z{=}0$};
    \node at (2.5,4.55)
        {\Large $h_{\rm m}\!\int_{z=0}\partial_{z}\phi$};
\end{scope}

\draw[->] (7.7,2) -- (6.1,2);
\node at (6.9,2.35) {Resolve};

\begin{scope}[shift={(2.5,-6)}]
    \fill[gnyfill] (0,0) rectangle (2.8,3.7);
    \fill[shadefill] (2.8,0) rectangle (3.2,3.7);
    \draw[guide] (2.8,0)--(2.8,3.7);

    \foreach \y in {0.3,0.65,...,3.3}
        \draw[hatch] (2.9,\y)--(3.1,\y+0.18);

    \node at (1.4,2) {\Large GNY};
    \node at (3.6,2) {\Large $|B_1\rangle$};
\end{scope}

\begin{scope}[shift={(7.8,-6)}]
    \fill[shadefill] (0,0) rectangle (0.4,3.7);
    \fill[gnyfill] (0.4,0) rectangle (3.2,3.7);
    \draw[guide] (0.4,0)--(0.4,3.7);

    \foreach \y in {0.3,0.65,...,3.3}
        \draw[hatch] (0.1,\y)--(0.3,\y+0.18);

    \node at (-.4,2) {\Large $\langle B_1|$};
    \node at (1.8,2) {\Large GNY};
\end{scope}

\draw[guide,dashed] (6.75,-6)--(6.75,-2.3);
\node at (6.75,-2) {\Large $ {\chi}^I$};
\node at (6.75,-6.2) {$z{=}0$};

\draw[->] (3,-0.6) -- (5.5,-1.8);
\node[right] at (1.75,-1.25) {Strip RG flow};

\draw[-> ] (10.6,-.6) -- (8.1,-1.8);
\node[right] at (10.,-1.25) {Defect RG flow};

\end{tikzpicture}
}
    \caption{The extraordinary surface defect viewed as a squeezed limit of the domain wall $m(z)$ for the fermion mass (equivalently the pseudoscalar $\phi$) in a strip. Here $m_*>0$ and $h_{\rm m}>0$. }
    \label{fig:squeeze}
\end{figure}

The emergence of (anti)chiral fermions in \eqref{mainsol} admits a direct explanation from the anomaly considerations above, together with the observation that the extraordinary surface defect can be realized by squeezing a domain wall profile $m(z)$ onto a thin strip, as illustrated in Fig.~\ref{fig:squeeze}. The domain wall coupling in the GNY CFT can be written equivalently as $m(z)\phi$, and it follows from \eqref{dw} that in the squeezing limit $\ell \ll L \ll z_{\rm typ}$, with $z_{\rm typ}$ the characteristic transverse scale set by operator insertions, one recovers the gradient deformation defining the extraordinary surface in \eqref{mainsol}.

On the other hand, if one first takes the limit $m_* L \gg m_* \ell \gg 1$, the RG flow on the strip with the domain wall mass profile implies that the bulk of the strip is gapped, leaving behind $N$ chiral fermions localized at the surface $z=0$. From the viewpoint of the region to the left of the strip, the boundary condition $|B_1\ra$ is selected, as it is stable under boundary deformations by $\phi$. Similarly, the right region imposes the boundary condition $\la B_1|$. Assuming that the squeezing limit commutes with the RG flow on the strip, we are thus led to the proposal in \eqref{mainsol}.

Since the extraordinary surface is connected to the trivial surface by a local deformation that respects the surface Poincar\'e symmetry, the gravitational anomaly for the extraordinary surface in the IR must be zero. Therefore the factorized interface $|B_1\ra\la B_1|$ must supply chiral central charge $c_L-c_R=-N/2$. One way to see this is to study the boundary gravitational anomaly of the theory on a strip $z\in [0,L]$ between $\la B_1|$ and $|B_1\ra$. Here the fermions satisfy boundary conditions $\gamma^1\Psi=-\Psi$ at $z=0,L$. By deforming the bulk theory to free fermions and taking limit of small $L$ (compared to longitudinal directions), one finds that the zero modes are described by $N$ $2d$ anti-chiral Majorana fermions \footnote{Any $3d$ spin QFT is defined up to a quantized Chern-Simons counter-term $k{\rm CS}_g$ with $k\in \mZ$ which modifies the contact terms in the stress-tensor two-point functions and shifts the boundary gravitational anomaly \cite{Closset:2012vg,Closset:2012vp,Prochazka:2019bhv,Wang:2020xkc} but this effect is canceled between one boundary $|B\ra$ and its transverse reflection $\la B |$.}. Another way to see this is to realize the normal boundary as a limit of a domain wall profile modified from \eqref{dw} so that in the thin-wall limit $\rho(m(z))=1-\theta(z)/2$ where $\theta(\cdot)$ is the Heaviside step function \footnote{Here the constant shift in $\rho$ is fixed by requiring $\rho(0)=1/2$ so that the anomaly theory \eqref{manomaly} is $T$-invariant \cite{Cordova:2019uob}. This fixes the Chern-Simons counterterm in the previous footnote.}. The gravitational anomaly of $|B_1\ra$ then follows from inflow as around \eqref{manomaly}.

\section{Solving the Extraordinary Surfaces at Large N} \label{sec:largeN}
The critical points of the GNY and the Gross–Neveu (GN) models lie in the same universality class at large $N$, and can be efficiently analyzed using the effective action
\begin{align}\label{GNbulk} S_{ \star} = -\frac{1}{2}\int dz d^2\bold{x} \, \bar\Psi^I(\slashed{\partial}+\phi)\Psi^I\,, \end{align}
which arises from a Hubbard–Stratonovich (HS) transformation of the GN model and provides a convenient starting point for systematic $1/N$ expansions \cite{Zinn-Justin:1991ksq}. The HS scalar is identified, up to a normalization constant, with the pseudoscalar in the GNY model, and will also be denoted by $\phi$. Throughout this section, we work in Euclidean signature. The corresponding spinor conventions are summarized in Appendix~\ref{spinorcon}.

The extraordinary surface defect is defined by the defect Lagrangian $L_{\rm ext} = h_{\rm m}\partial_z \phi\delta(z)$ in \eqref{LDefect}, whose RG flow we analyze at large $N$. As explained in Appendix~\ref{powercounting}, this flow lies beyond the regime of standard conformal perturbation theory.
Instead, we follow the strategy of \cite{Krishnan:2023cff} together with the approach of \cite{Giombi:2020rmc,Giombi:2021cnr}. The same procedure applies to the normal surface defect defined by $L_{\rm norm} = h_{\rm o}\phi\delta(z)$. The corresponding RG flow is analyzed in Appendix~\ref{app:norm} with the exact IR fixed points summarized in \eqref{normsol}.

The first step is to map the defect problem to a boundary problem via the folding trick. We introduce two copies of the fermion $\Psi_{1,2}$, and the pseudoscalar $\phi_{1,2}$, defined on the half-space $z \geq 0$, obtained by folding the original fields on $z \in \mR$ across the surface at $z=0$ using the transverse reflection,
\ie 
\begin{cases}
\Psi^I(z,\bold{x})=\Psi^I_1(z,\bold{x})\,, \\  
        \Psi^{I}(-z,\bold{x})=\gamma^{1}\Psi_{2}^I(z,\bold{x})\,,
\end{cases}
\begin{cases}
         \phi(z,\bold{x})=\phi_1(z,\bold{x})\,, \\  \phi(-z,\bold{x})=-\phi_2(z,\bold{x})\,.
\end{cases}
\label{psi1psi2}
\fe 
The action \eqref{GNbulk} becomes
\begin{gather}\label{S*1}
    S_{ \star}=-\frac{1}{2}\sum_{a=1}^2\int_{z \geq 0} dz d^2 \bold{x}   \bar\Psi^I_a\left[\slashed{\partial} +  \phi_a(z) \right]\Psi^I_a\,.
\end{gather}
The transverse parity \eqref{parity} is equivalent to swapping the two species of fields. The most general O$(N)$ invariant boundary condition for the fermions is parameterized by an angle $\vartheta$
\ie\label{generalbc}
   & \left.\gamma^1\Psi^I_1 \right|_{z=0} =\left. \sin(\vartheta)\Psi^I_1 +\cos(\vartheta)\Psi^I_2 \right|_{z=0} \,,
   \\
   & \left.\gamma^1\Psi^I_2 \right|_{z=0} =\left. \cos(\vartheta)\Psi^I_1 -\sin(\vartheta)\Psi^I_2 \right|_{z=0} \,.
\fe 
In the following, we solve the extraordinary surface explicitly by determining the fermion and pseudoscalar propagators in the presence of the boundary obtained via folding along the surface. In fact, we find a one-parameter family of conformal extraordinary surfaces in the large $N$ limit. This is because to leading order in $N$, the pseudoscalar has dimension $\Delta=1$ and thus $L_{\rm ext}$ is marginal and in fact exactly marginal \footnote{In contrast, the deformation $h\phi^2\D(z)$ is marginal but not exactly marginal in the large $N$ limit \cite{Ord}. See Appendix~\ref{powercounting} for relevant three-point functions.}.
The solutions in \eqref{mainsol} arise in a degeneration limit and are precisely those that persist after including ${1/N}$ corrections.

\subsection{The fermion propagators}
The pseudoscalar acquires a nonzero one-point function $\langle\phi\rangle$ due to the surface deformation $L_{\rm ext}$. In the large $N$ limit,  this is fixed  by the bulk scaling dimension of $\phi$ and transverse parity, up to an overall coefficient which we denote by $\mu$
\begin{align}\label{sigma1pt}
    \langle\phi(z,\bold{x})\rangle = \frac{\mu}{z}\,, \quad z\in\mathbb R\,.
\end{align}
Because $\mu>0$ and $\mu<0$ are related by time-reversal (longitudinal parity), it suffices to work with $\mu\ge 0$ below.

In the boundary formulation upon folding, \eqref{sigma1pt} becomes \ie 
\langle\phi_1(z,\bold{x})\rangle=\langle\phi_2(z,\bold{x})\rangle = \frac{\mu}{z}\,, \quad z >0
\label{phisaddle}
\fe Following \cite{Giombi:2020rmc,Giombi:2021cnr}, we absorb the $z$ dependence by mapping the flat half-space to hyperbolic space $\mathbb{H}_3$ wth metric $ds^2=\frac{dz^2+d\bold{x}^2}{z^2}$ and implement the Weyl transformations  $\Psi_a\to z^{-1}\,\Psi_a\,, \phi_a\to z^{-1}\,\phi_a$ \footnote{The fermion and pseudoscalar fields all have dimension $\Delta=1$ in the large $N$ limit. The ${1\over N}$ anomalous dimensions is positive for the fermions and negative for the pseudoscalar. \cite{Gracey:1990wi, Gracey:1993kc}}. The resulting action in $\mathbb{H}_3$ reads
\begin{gather}
    S_{\star}=-\frac{1}{2}\sum_{a=1}^2\int_{z\ge 0}dzd^2\bold{x} \sqrt{g} \,  \bar\Psi^I_a\left[\slashed{\nabla} +  \phi_a \right]\Psi^I_a\,,
    \label{H3action}
\end{gather}
where $\sqrt{g} = \frac{1}{z^3}$, and  $\slashed{\nabla}= z \slashed{\partial}-\gamma^1$ is the Dirac operator in the Poincar\'e coordinates of $\mathbb{H}_3$. For all $\mu$, we will find that \eqref{phisaddle} is a solution to the saddle point equation from \eqref{H3action} subject to particular boundary conditions for the fermions. 

Given the saddle point of $\phi_a$, the fermion propagators $\langle\Psi^I_a(x_1)\bar\Psi^J_b(x_2)\rangle\equiv \delta^{IJ}G_{ab}(x_1, x_2)$ at large $N$  satisfy 
\begin{align}
    (\slashed{\nabla}+\mu) G_{ab} (x_1, x_2) = -\delta_{ab} \delta^3(x_1, x_2)\,.
    \label{DEq}
\end{align}
Away from the coincident limit $x_1=x_2$, we solve \eqref{DEq} using the following ans\"atz for $G$ \cite{Mueck:1999efk, Basu:2006ti}
\begin{align}\label{Gansatz}
    G(x_1, x_2) = \frac{1}{\sqrt{2z_1 z_2}}\left(\slashed{x}_{12}\frac{\beta(v)}{\sqrt{v-1}}-\gamma^1\slashed{\bar x}_{12}\frac{\alpha(v)}{\sqrt{v+1}} \right)\,,
\end{align}
where $ x_{12}\equiv  (z_1-z_2, \bold{x}_{12})\,, \bar x_{12} \equiv (-z_1-z_2, \bold{x}_{12})$ and 
\begin{align}\label{xxv}
     v\equiv \frac{z_1^2+z_2^2+\bold{x}^2_{12}}{2z_1 z_2}\,.
\end{align} 
In the embedding space coordinates of $\mathbb{H}_3$ \eqref{Pcoord}, the conformal cross-ratio $v=-X_1\cdot X_2$ coincides with  the usual $\mathbb{H}_3$ invariant.
Acting with the Dirac operator $\slashed{\nabla}+\mu$ on $G$ yields two coupled differential equations for $\alpha(v)$ and $\beta(v)$, which admit two linearly independent solutions \footnote{The two solutions $G_\pm$ in \eqref{Gpmexp} are degenerate at $\mu=1/2$ where the fermionic Breitenlohner-Freedman (BF)  bound in AdS$_3$ is saturated. A new logarithmic solution emerges at this point and is given by $\left.\partial_\mu (G_+- G_-)\right|_{\mu=1/2}$. While its leading fall-off at large $r$ corresponds to a $\hat\Delta=1/2$ boundary fermion, this solution is not normalizable \cite{Amsel:2008iz}. This is another indication that the $\hat\Delta=1/2$ fermion is completely localized on the boundary and described by a singleton \cite{Nilsson:2018lof}.} 
\begin{align}\label{Gpmexp}
    G_\pm = \frac{{\rm csch}(r) e^{\mp \mu r}}{2\sqrt{z_1 z_2}}\left(\slashed{x}_{12}\!\frac{\coth({\frac{r}{2}})\!\pm\! 2\mu}{\sinh(\frac{r}{2})}\!-\!\gamma^1\slashed{\bar x}_{12}\frac{2\mu\!\pm\!\tanh(\frac{r}{2})}{\cosh(\frac{r}{2})}\right)\,,
\end{align}
where $r\equiv {\rm arccosh}(v)$ and
$G_\pm$ satisfy the boundary condition
\begin{align}\label{gammaG}
    \left.\gamma^1 G_\pm (x_1, x_2)\right|_{z_1\to 0} = \mp  \left.G_\pm (x_1, x_2)\right|_{z_1\to 0}\,.
\end{align}

The fermion propagators $G_{11}$ and $G_{12}$ can be expressed as linear combinations of $G_\pm$, with coefficients fixed by the bulk conformal block expansion. According to \cite{Herzog:2022jlx}, for a bulk operator of dimension $\Delta$, there are two channels of spinor bulk conformal blocks, corresponding to scalar primaries of opposite parity in the bulk OPE limit.

Let $G_{11} = c_+ G_+ + c_- G_-$. For generic coefficients $c_\pm$, both parity channels at dimension $2$ contribute to $G_{11}$. However, at $N=\infty$, the $\Psi \times \Psi$ OPE contains a unique scalar operator of dimension $2$, which is parity-even. To eliminate the parity-odd contribution, we must impose $c_+ = c_-$. The overall normalization is fixed by the identity channel: the most singular term in the bulk OPE limit should match, up to a Weyl transformation, the massless fermion propagator in $\mathbb{R}^3$ \footnote{In our normalization, this free fermion propagator is $  
-\frac{\slashed{x}_{12}}{4\pi |x_{12}|^3}
$.}.

The propagator $G_{12}$ is non-singular in the bulk OPE limit, since in the original interface setup it involves fermions located on opposite sides of the planar defect. Moreover, it must exhibit the same large-$v$ behavior as $G_{11}$, because $\Psi_1$ and $\Psi_2$ are governed by the same leading operator in the boundary OPE channel. The first condition implies $G_{12} \propto G_+ - G_-$, while the second fixes the overall normalization.
Altogether, the solution to $G_{ab}$ is
\begin{align}\label{12pm}
    G_{ab}=-\frac{1}{16\pi}\times\begin{cases}
         G_++ G_-\,, & a=b\,,\\
         -G_++G_-\,,&a\not=b\,.
    \end{cases}
\end{align}
Combining \eqref{gammaG} and \eqref{12pm} implies  
\begin{align}\label{gG12}
       \left.\gamma^1 G_{1,a} (x_1, x_2)\right|_{z_1\to 0} =   \left.G_{2,a} (x_1, x_2)\right|_{z_1\to 0}\,,
\end{align}
which corresponds to $\vartheta=0$ in \eqref{generalbc}.

We define transverse parity-even fermions $\Psi_{\rm e}^I =\frac{\Psi^I_1+\Psi_2^I}{\sqrt{2}}$ and parity-odd fermions $\Psi_{\rm o}^I =\frac{\Psi^I_1-\Psi_2^I}{\sqrt{2}}$. Their two-point functions decouple at large $N$, i.e. $\left\langle \Psi^I_{\rm e} \bar\Psi_{\rm o}^J\right\rangle = 0$. Within each parity sector, the two-point functions are
\begin{align}
    \left\langle \Psi^I_{\rm e}  \bar\Psi^J_{\rm e}\right\rangle = - \delta^{IJ}\frac{G_-}{8\pi}\,, \quad  \left\langle \Psi^I_{\rm o}  \bar\Psi^J_{\rm o}\right\rangle = - \delta^{IJ}\frac{G_+}{8\pi}\,.
\end{align}
We  read off the boundary dimensions of $\Psi^I_{\rm e/o}$ from the large $r$ (boundary OPE limit) behavior of $G_\pm$: 
\begin{equation}\label{fermlargeNanom}
\begin{split}
    &\Psi^I_{\rm e}:  (h, \bar h)=\left(\frac{3}{4}-\frac{\mu}{2},\frac{1}{4}-\frac{\mu}{2} \right)\,, \quad \hat\Delta_{\rm e} = 1-\mu\,, \\
    &\Psi^I_{\rm o}:  (h, \bar h)= \left(\frac{1}{4}+\frac{\mu}{2},\frac{3}{4}+\frac{\mu}{2} \right) \,, \quad \hat\Delta_{\rm o} = 1+\mu \,.
    \end{split} 
\end{equation}
 The unitarity bound of fermions in $2d$ imposes $\mu\le \frac{1}{2}$. 

For a free Majorana spinor in $\mathbb{H}_3$ of boundary dimension $\hat\Delta$, its one-loop free energy is \cite{Giombi:2021cnr}
\begin{align}\label{FD}
    F_M(\hat{\Delta}) = \frac{V_3}{48\pi}(\hat{\Delta}-1)\left(4(\hat{\Delta}-1)^2-3\right)\,, 
\end{align}
where $V_3=-2\pi\log (R)$ is the regularized volume of $\mathbb{H}_3$ \eqref{AdSvol}. At order $N$, the defect $b$-anomaly is proportional to the coefficient of $\log(R)$ in $F_M(\hat{\Delta}_{\rm e})+F_M(\hat{\Delta}_{\rm o})$, which vanishes identically for arbitrary $\mu$, justifying our assumption on the saddle below \eqref{H3action} and
indicating the existence of  a line of conformal defects parameterized by $-\frac{1}{2}\le \mu\le \frac{1}{2}$.

\subsection{The pseudoscalar propagator}
We now turn to study the two-point function of the pseudoscalar $\phi$.
Let $\delta\phi$ be the fluctuation of $\phi$ around its saddle point \eqref{sigma1pt}.  In the original defect set-up, the  two-point function of $\delta\phi$ is subject to the Schwinger-Dyson equation,
\begin{align}\label{phiphi}
    \frac{1}{2}\int d^3 x_2\, &\tr\left(\langle \Psi^I(x_1)\bar\Psi^J(x_2)\rangle\langle \Psi^J(x_2)\bar\Psi^I(x_1)\rangle\right)\nonumber\\
    &\times \langle\delta\phi(x_2)\delta\phi(x_3)\rangle = \delta^3(x_1,x_3)\,.
\end{align}
In Appendix~\ref{app1}, we solve this by mapping \eqref{phiphi} to $\mathbb{H}_3$ and using the spectral method there. Here we present the results. For the transverse parity-even combination $\delta\phi_{\rm e} = \frac{\delta\phi_1+\delta\phi_2}{\sqrt{2}}$, the two-point function reads  
\begin{align}\label{H+expm}
    &H_{\rm e} (r)=\frac{32}{N\pi}\sum_{n\ge 1}\frac{n^2(1-(-)^{n}\cos(2\pi\mu))}{n^2-4\mu^2} G_n(r)\,,
    \end{align}
where $G_s(r)=\frac{e^{-s r}}{4\pi \sinh(r)}$ is the scalar boundary conformal block with 
dimension $\hat\Delta = 1+s$.  From the boundary block expansion \eqref{H+expm}, we find operators of dimension $\hat\Delta = 2, 3, 4,\cdots$ in the boundary OPE of  $\delta\phi_{\rm e}$. In particular, the leading operator with dimension $2$, denoted by $\hat\CO(\mu)$,  tunes the boundary along the line of fixed points. Note that all even dimensional boundary operators decouples at $\mu=\frac{1}{2}$, highlighting the degenerate nature of this point. 

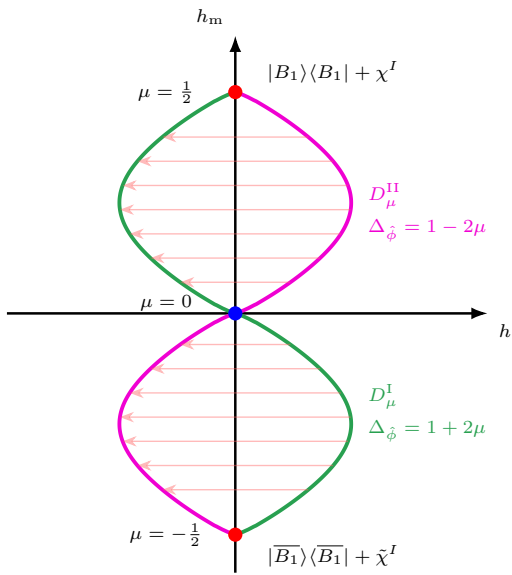
\begin{figure}[!htb]
    \centering
\begin{tikzpicture}[
  x=0.92cm,y=0.92cm,
  >=Latex,
  every node/.style={font=\scriptsize},
  axis/.style={line width=0.95pt,-{Latex[length=2.2mm,width=1.6mm]}},
  loopline/.style={line width=1.45pt,line cap=round,line join=round},
  flow/.style={red!60,opacity=0.45,line width=0.52pt,-Stealth},
  dot/.style={circle,fill=black,inner sep=1.9pt}
]
\definecolor{myblue}{RGB}{240,0,210}
\definecolor{mymagenta}{RGB}{40,160,80}

\def\yt{3.20}
\def\a{1.58}
\def\p{0.82}
\pgfmathdeclarefunction{fw}{1}{%
  \pgfmathparse{\a*(sin(180*#1/\yt))^\p * (1 + 0.06*sin(180*#1/\yt)^2)}%
}

\draw[axis] (-3.30,0) -- (3.65,0) node[below right=1pt] {$h$};
\draw[axis] (0,-3.75) -- (0,4.02) node[above left=1pt] {$h_{\rm m}$};

\draw[loopline,mymagenta]
  plot[domain=0:\yt,samples=220,variable=\t] ({-fw(\t)},{\t});
\draw[loopline,myblue]
  plot[domain=\yt:0,samples=220,variable=\t] ({ fw(\t)},{\t});

\draw[loopline,myblue]
  plot[domain=0:-\yt,samples=220,variable=\t] ({-fw(-\t)},{\t});
\draw[loopline,mymagenta]
  plot[domain=-\yt:0,samples=220,variable=\t] ({ fw(-\t)},{\t});

\node[dot,red] at (0,\yt) {};
\node[dot,red] at (0,-\yt) {};
\node[dot,blue] at (0,0) {};

\foreach \yy in {2.55,2.20,1.85,1.50,1.15,0.80,0.45,-0.45,-0.80,-1.15,-1.50,-1.85,-2.20,-2.55}{
  \pgfmathsetmacro{\xx}{fw(abs(\yy))}
  \draw[flow] (\xx,\yy) -- (-\xx,\yy);
}

\node[anchor=west] at (0.34,3.50) {$|B_1\rangle\langle B_1|+\chi^I$};
\node[anchor=west] at (0.34,-3.50) {$|\overline{B_1}\rangle\langle \overline{B_1}|+\tilde \chi^I$};

\node[text=myblue,anchor=west] at (1.8,1.7) {$D_\m^{\mathrm{II}}$};
\node[text=myblue,anchor=west] at (1.8,1.2) {$\Delta_{\hat\phi}=1-2\mu$};
\node[text=mymagenta,anchor=west] at (1.8,-1.2) {$D_\m^{\mathrm{I}}$};
\node[text=mymagenta,anchor=west] at (1.8,-1.7) {$\Delta_{\hat\phi}=1+2\mu$};

\node[] at (-1,\yt) {$\m={1\over 2}$};
\node[] at (-1,0.15) {$\m=0$};
\node[] at (-1,-\yt) {$\m=-{1\over 2}$};

\end{tikzpicture}
   \caption{Emergent conformal manifold of extraordinary surface defects at large $N$. The two branches, $D_\mu^{\rm I}$ and $D_\mu^{\rm II}$, parametrized by $\mu\in[-1/2,1/2]$, are shown in  green and pink, respectively. The pink horizontal arrows represent double trace flows between the two branches (at subleading order in $1/N$). The solid dots persist as fixed points at finite $N$: the red points are stable (extraordinary conformal surfaces), while the blue point is unstable (the trivial surface). Here we use $h,h_{\rm m}$ to denote the renormalized dimensionless couplings that preserve the transverse parity. The ``kinks'' at the tips in the diagram are to emphasize degenerate features of the conformal manifold there, where conformal multiplets undergo recombination and decoupling.}
    \label{fig:eight}
\end{figure}

For the parity-odd combination $\delta\phi_{\rm o} = \frac{\delta\phi_1-\delta\phi_2}{\sqrt{2}}$, there are two choices for the bulk two-point function $\langle \delta\phi_{\rm o}\delta\phi_{\rm o}\rangle$, which differ only in the leading boundary OPE channel,
\begin{equation}\label{H-expm}
    \begin{split}
    &H^{\rm I}_{\rm o} (r)=\frac{32\mu \, G_{2\mu}(r)}{N \tan(\pi\mu)} \!+\! \frac{64}{N\pi}\sum_{n\ge 1}\frac{n^2}{ n^2\!-\!\mu^2}G_{2n}(r)\,,\\
     &H^{\rm II}_{\rm o} (r)=\frac{32\mu \, G_{-2\mu}(r)}{N \tan(\pi\mu)} \!+\! \frac{64}{N\pi}\sum_{n\ge 1}\frac{n^2}{n^2\!-\!\mu^2}G_{2n}(r)\,.
     \end{split}
\end{equation}
Combined with the common two-point function $H_{\rm e}$ in the even sector, $H^{\rm I}_{\rm o} (r)$ and $H^{\rm II}_{\rm o} (r)$ 
define two different families of BCFTs for the folded theory \eqref{H3action}. We have thus discovered two families of conformal extraordinary surfaces at the leading order of large $N$ expansion parametrized by $\mu$. We label the two families by $D^{\rm I}_\mu$ and $D^{\rm II}_\mu$ respectively as illustrated in Fig.~\ref{fig:eight}.

For generic $\mu$, the leading parity-odd scalar operator has dimension $1+2\mu$ along the branch $D^{\rm I}_\mu$ and dimension $1-2\mu$ along the branch $D^{\rm II}_\mu$. The subleading boundary operators, with dimension $\hat\Delta=3,5,7, \cdots$, exist on both branches.
There is an RG flow from $D^{\rm II}_\mu$ to $D^{\rm I}_\mu$, triggered by the double trace deformation of the $\hat\Delta=1-2\mu$ operator. At $\mu=\frac{1}{2}$, the $\hat\Delta=1\pm 2\mu$ operator decouples from the boundary OPE of $\delta\phi_{\rm o}$, and the two branches join, producing a one-dimensional conformal manifold of nontrivial topology. We will discuss features of this conformal manifold in Section~\ref{sec:defspace}.

By using the two-point function of the pseudoscalars, we show in Appendix \ref{app2} that the continuous families of defect fixed points at $N=\infty$ are lifted to three points $\mu=0, \pm \frac{1}{2}$ by the $1/N$ corrections. We will find that the $\mu=\pm {1\over 2}$ points precisely correspond to the RHS of \eqref{mainsol}.

\section{Traversing the Coupling Space: Topology, Distance and Degenerations} 
\label{sec:defspace}
In the previous section, we found that at $N=\infty$, for a given $\mu\in(-\frac{1}{2},\frac{1}{2})$ parameterizing the bulk one-point function of $\phi$, there exist two distinct BCFTs, distinguished by the dimension of the leading parity-odd boundary scalar. The two branches of defect fixed points join at $\mu=0, \pm\frac{1}{2}$, giving rise to an emergent defect conformal manifold with the topology of the figure ``8'',  as shown in Figure \ref{fig:eight}. The origin at $\mu=0$ corresponds to the trivial defect. The behavior of the conformal manifold in its vicinity, together with the location of the branches $D^{\rm I}_{\mu}$ and $D^{\rm II}_{\mu}$, is analyzed in Appendix~\ref{powercounting} (see in particular the discussion around \eqref{betaCross}).  Below we discuss the $\mu=\frac{1}{2}$ point in detail which exhibits interesting degeneration phenomena (generalizing those in $2d$ CFT \cite{Kontsevich:2000yf,Roggenkamp:2003qp,Ooguri:2024ofs,Soibelman:2025enm}).

\subsection{Degenerations at the tips of ``8''}
As $\mu$ approaches $\frac{1}{2}$, the long multiplet of dimension
$\left(\frac{3}{4}-\frac{\mu}{2},\frac{1}{4}-\frac{\mu}{2} \right) $ furnished by the leading transverse parity-even boundary fermion  splits into a short multiplet $(\frac{1}{2}, 0)$ and a long multiplet $(\frac{1}{2},1)$. The short multiplet describes a  chiral Majorana fermion in $2d$, and the long multiplet $(\frac{1}{2},1)$ is the same as the representation carried by the leading parity-odd fermion at the normal boundary $|B_1\ra$ \cite{Giombi:2021cnr}. This identifies the $\mu = \frac{1}{2}$ point with the factorized interface with emergent $2d$ chiral fermions in \eqref{mainsol}.
  
An immediate consistency check of this proposal is the $b$-anomaly at large $N$. On the one hand, the $\mathbb{H}_3$ free energy of 2$N$  Majorana spinor with boundary dimension $\hat\Delta = \frac{3}{2}$ is $2N F_M(\frac{3}{2}) = \frac{N}{12}\log(R)$ from \eqref{FD}.
So the  $b$-anomaly of the factorized defect $|B_1\rangle\langle B_1|$ is $b_{|B_1\rangle\langle B_1|} = -\frac{N}{4}$. On the other hand, the $b$-anomaly of a single 2$d$ chiral Majorana fermion is $\frac{c_L+c_R}{2} = \frac{1}{4}$. Therefore, the combined system $|B_1\rangle\langle B_1|+\chi^I$ has a vanishing $b$-anomaly at order $N$, which is the same as other points on the conformal manifold. In Appendix~\ref{app2}, we find that the order $N^0$ contribution in $b_{|B_1\rangle\langle B_1|} $ is equal to $ -\frac{9}{8}$. So the proposal \eqref{mainsol} obeys the defect $b$-theorem $b_{\rm UV}>b_{\rm IR}$ at the $N^0$ order since $b_{\rm UV}= 0$.

Let $\psi^I_L$ and $\psi^I_R$ be the dimension $\frac{3}{2}$ boundary fermions in the two decouple $|B_1\ra$ boundaries. We identify the marginal operator that decouples from $\phi_{\rm e}$ at $\mu=\frac{1}{2}$ as 
\begin{align}\label{Oe}
    \CO_{\rm e} = \frac{1}{\sqrt{2N}}(\psi_L^I+\psi^I_R)\chi^I\,.
\end{align} 
Similarly, the marginal operator that decouples from $\phi_{\rm o}$ at $\mu=\frac{1}{2}$ is identified as $\frac{1}{\sqrt{2N}}(\psi_L^I-\psi^I_R)\chi^I$. We can probe the neighborhood of $\mu=\frac{1}{2}$ by turning on the perturbation $s\int d^2 \bold{x} \,\CO_{\rm e}$ (see Appendix \ref{ConfPT} for details and curious phenomena).  The anomalous dimension of $\chi^I$ induced by this perturbation is determined to be $\Delta_\chi = \frac{1}{2}+\pi^2 s^2/N$ \eqref{CPTchiral}. By identifying with the  dimension $1-\mu$ of the boundary fermion corresponding to  $\Psi^I_{\rm o}$, this gives at small $s/\sqrt{N}$ \cite{myfootnote},
\begin{align}\label{smu}
   \frac{ \pi^2 s^2}{N} = \frac{1}{2}-\mu +O\left(\left(\mu-\frac{1}{2}\right)^2\right)\,.
\end{align} 
This allows us to deduce the beta function of $\mu$ in terms of that of $s$. The $1/N$ correction to the dimension of $\psi^I_{L/R}$ is $\gamma_\psi=\frac{1}{2N}$ \cite{DGS} and thus $\beta_s = \frac{1}{2N} s+O (s^3)$. Consequently,
\begin{align}
    \beta_\mu = -\frac{2\pi^2 s}{N}\beta_s = \frac{1}{N}\left(\mu-\frac{1}{2}\right)+O\left(\left(\mu-\frac{1}{2}\right)^3\right)\,,
\end{align}
which implies that $\m={1\over 2}$ is a stable fixed point.

Following the same multiplet recombination argument, we also find a decoupled massless parity-odd scalar $\varphi$, when approaching the $\mu=\frac{1}{2}$ point along the $D^{\rm II}_\mu$ branch. This emergent massless scalar is consistent with the double-trace deformation between the two branches.
According to \eqref{Fss}, for any $0<\mu<\frac{1}{2}$, the difference of free energy between the two branches is $\delta F= -\frac{(2\mu)^3}{3}\log(R)$. At $\mu=\frac{1}{2}$, it corresponds to $\delta b=1$, which is the same as the central charge of a real massless scalar \footnote{Unlike the emergent chiral fermions, this scalar is not expected to be an intrinsic feature of the extraordinary fixed point at finite $N$ (e.g. from anomaly considerations). We will not attempt to give a complete account of its origin at finite $N$ here. 
Our expectation is that, at finite $N$, the tip of the branch $D_\m^{\rm II}$ at $\m=1/2$ is instead described by a factorized interface involving the special boundary condition $|B_2'\ra$. A suggestive hint in this direction is that the pseudoscalar operator acquires vanishing scaling dimension if one naively extrapolates the large $N$ analysis of \cite{Giombi:2021cnr} to $d=3$.}.

The degeneration phenomena described here at large $N$, from the $\mH_3$ perspective, are closely related to Higgsing/deHiggsing transitions in AdS/CFT \cite{Nilsson:2018lof,Duff:2025tot}, in which the masses of bulk AdS fields are tuned by squashing the internal compact geometry. The emergent boundary chiral fermion and massless scalar are realized as singletons, rather than propagating bulk AdS fields \cite{Dirac:1963ta,Flato:1990eu,Flato:1999yp}, and play a crucial role in ensuring the continuity of these transitions \footnote{See also \cite{Ohl:2012bk}, where it is shown explicitly how a singleton mode confined to the boundary emerges from a bulk AdS field upon saturating the unitarity bound.}.

\subsection{Zamolodchikov metric  and Zamolodchikov norm}
A natural metric in the couplings space is given by the Zamolodchikov metric \cite{Zamolodchikov:1986gt}, which is the two-point function coefficient of the marginal operator (see e.g. \cite{Ooguri:2024ofs} for a recent review).
To determine the Zamolodchikov metric $C(\mu)$ on the $1d$ conformal manifold in Figure~\ref{fig:eight}, we first note that $\mu$ conjugates to a boundary operator of the form $\kappa(\mu)\hat\CO_\mu$, where $\hat\CO_\mu$ is the (canonically normalized) marginal operator in the boundary OPE  of $\delta\phi_{\rm e}$. The corresponding OPE  coefficient can be extracted from \eqref{H+expm} 
\begin{align}
a^2_{\hat\CO_\mu}=\frac{32\cos^2(\pi\mu)}{\pi^2 N(1-4\mu^2)}\,.
\end{align}   
The factor $\kappa(\mu)$ is then fixed by the relation 
\begin{align}
    \partial_\mu\langle\phi_{\rm e}(z)\rangle = -\kappa(\mu)\int d^2\bold{x}\langle\delta\phi_{\rm e}(z) \hat\CO_\mu(\bold{x})\rangle\,,
\end{align}
where $\langle\phi_{\rm e}(z)\rangle =\sqrt{2}\mu$ on $\mathbb{H}_3$. The integral on the RHS gives $\pi a_{\hat\CO_\mu}$, and hence 
\begin{align}
    C(\mu) = \kappa^2(\mu) =N \frac{1-4\mu^2}{(4\cos(\pi\mu))^2}\,.
\end{align}
The Zamolodchikov distance between $\mu=0$ and $\mu=\frac{1}{2}$ is
\begin{align}
    L =\int_0^{\frac{1}{2}}d\mu \sqrt{C(\mu)} \approx 0.22 \sqrt{N}\,.
\end{align}
Note that this is consistent with the general argument in \cite{Ooguri:2024ofs} that degenerations on the $2d$ conformal manifold are located at infinite Zamolodchikov distance because the defect conformal manifold here only exists at infinite $N$.

The parity-odd dimension $3$ operator in the boundary OPE of $\delta\phi_{\rm o}$ corresponds to the displacement operator $\mathbb D$ of the interface, which is defined by  \cite{Billo:2016cpy}
\begin{align}\label{Ddef}
   \partial_\mu T^{\mu z} = - \mathbb D \,\delta(z) \,.
\end{align} 
The displacement operator arises from the breaking of translation invariance in the $z$ direction by the interface. According to the definition \eqref{Ddef}, the normalization of $\mathbb D$ is fixed by the canonically normalized stress energy tensor $T_{\mu\nu}$. The two-point function coefficient of $\mathbb D$, known as the  Zamolodchikov norm $C_{\mathbb D}$, is thus a physical quantity which is non-negative by unitarity and vanishes only if defect is topological \cite{Billo:2016cpy}. We compute the Zamolodchikov norm as a function of $\mu$, by using the boundary block expansion \eqref{H-exp}.
First, \eqref{Ddef} implies a fundamental Ward identity 
\begin{align}
    a_{\cO}\Delta_\cO = \frac{\pi}{2} b_{\cO\mathbb D}\,,
\end{align}
for any canonically normalized bulk primary operator $\cO$ with dimension $\Delta_\cO$, 
where $a_\cO$ is the one-point coefficient of $\cO$, i.e. $\langle \cO(z)\rangle= a_\cO/z^{\Delta_\cO}$, and $b_{\cO\mathbb D}$ is the coefficient of the mixed correlator $\langle \cO(z,\bold{x}_1 ) \mathbb{D}(\bold{x}_2)\rangle$ \cite{Billo:2016cpy}. In our case, we take $\cO=\CN_\phi^{-1} \phi$, where the coefficient $\CN_\phi$ is chosen such that $\cO$ is canonically normalized. Plugging this choice of $\cO$ into the Ward identity yields in the large $N$ limit,
\begin{align}
    b_{\cO\mathbb D} = \frac{2\mu}{\pi\,\CN_\phi}\,.
\end{align} 
 On the other hand, in the folded system,  $\langle \cO(z,\bold{x}_1 ) \mathbb{D}(\bold{x}_2)\rangle =\frac{1}{\sqrt{2}\CN_\phi}\langle  \phi_{\rm o}(z,\bold{x}_1 ) \mathbb{D}(\bold{x}_2)\rangle$. Let $k_{\mathbb D}$ be the boundary OPE coefficient of $\mathbb D$ in $\delta\phi_{\rm o}$, then $b_{\cO\mathbb D}=\frac{1}{\sqrt{2}\CN_\phi}k_{\mathbb D} C_{\mathbb D} $.  Combining this relation with the Ward identity, we find 
\begin{align}
    k_{\mathbb D}C_{\mathbb D} = \frac{2\sqrt{2}}{\pi}\mu\,.
\end{align}
In addition, given the boundary OPE coefficient $k_{\mathbb D}$, $H_{\rm o}^{\rm I,II}$ should include a term $k_{\mathbb D}^2 C_{\mathbb D} e^{-3 r}$ in the boundary OPE limit. Comparing with \eqref{H-exp}, we can identify $k_{\mathbb D}^2 C_{\mathbb D}$ with $\frac{32}{N\pi^2(1-\mu^2)}$. Altogether, we find the Zamolodchikov norm
\begin{align}
    C_{\mathbb D} = \frac{N}{4}\mu^2(1-\mu^2)\,.
\end{align}
In particular, it is non-negative and vanishes at $\mu=0$, as expected for the trivial defect which is topological.

\section*{Acknowledgments}
We thank Dean Carmi, Simone Giombi, Igor Klebanov, Zohar Komargodski, Hirosi Ooguri, David Poland, Fedor Popov, Massimo Porrati, Nathan Seiberg, Shu-Heng Shao for helpful discussions. We thank Max Metlitski for many helpful suggestions and critical discussions throughout the project.  
ZS is supported by the U.S. Department of Energy grant DE-SC0009988 and the Sivian Fund. The work of YW was supported in part by the BSF Grant 2024187 and by the Simons Junior Faculty Fellows program.

\bibliographystyle{apsrev4-1}
\bibliography{surfacedefect}

\newpage
\onecolumngrid

\appendix

\section{Spinor conventions}\label{spinorcon}
The coordinate system in (2+1)$d$ is $x^\mu=(t, z, x)$.
We use the following convention of gamma matrices 
\ie\label{gammaM}
\gamma_M^0 = i\tau^2\,,\quad \gamma_M^1 = \tau^3\,, \quad \gamma_M^2 = -\tau^1\,, 
\fe
where $\tau^i$ denotes the $i$-th Pauli matrix. They furnish a Majorana representation. The Dirac conjugate of a Majorana spinor $\Psi$ is given by
 $\bar\Psi =\Psi^t i\gamma^0_M = - \Psi^t\tau^2$.  The Dirac operator reads 
 \begin{align}
     \slashed{\partial}_M =\gamma_M^\mu\partial_\mu = \gamma_M^0\partial_t+ \gamma^1_M\partial_z+\gamma_M^2\partial_x\,.
 \end{align}
After performing the Wick rotation $t\to -i y$ to Euclidean signature, the Dirac operator becomes $ i\gamma^0_M \partial_y +    \gamma^1_M \partial_z+\gamma^2_M \partial_x$,
and hence it is natural to define the Euclidean gamma matrices
\ie\label{gammaE}
\gamma_E^3 = i\gamma_M^0=-\tau^2\,, \quad \gamma_E^1 = \gamma_M^1\,, \quad \gamma_E^2=\gamma_M^2\,.
\fe
The Wick-rotated spinors $\Psi^I$ are complex as the Euclidean spin group SU$(2)$ does not admit real two-dimensional representations. The Dirac conjugate takes the same form in Euclidean signature $\bar\Psi =- \Psi^t\tau^2$. We parameterize the coordinate in the Euclidean signature by $x^i=(z, \bold{x})\,, \bold{x}=(x, y)$.

\section{large-$N$ power counting of $\beta$-functions on the defect }
\label{powercounting}
In this appendix, we analyze the defect beta functions associated with the couplings in \eqref{LDefect} that preserve the transverse parity $P$ in \eqref{parity}. For convenience, we write the Euclidean action as
\begin{gather}\label{actionCPTorigin}
    S = -\frac{1}{2}\int dz \, d^2 \bold{x} \left( \bar{\Psi}^I \slashed{\partial} \Psi^{I} + \frac{1}{\sqrt{N}} \sigma \, \bar{\Psi}^I \Psi^{I} \right)
    - \hat{h}_0 \int d^2 \bold{x} \, \hat{\sigma}^2(\bold{x})
    - \hat{h}_{\rm m,0} \int d^2 \bold{x} \, \partial_{z}\hat{\sigma}(\bold{x}) \,,
\end{gather}
where $\hat{\sigma}(\bold{x}) \equiv \sigma(0,\bold{x})$ denotes the restriction of the bulk pseudoscalar to the defect. The couplings $\hat{h}_0$ and $\hat{h}_{\rm m,0}$ (proportional to $h$ and $h_{\rm m}$ in \eqref{LDefect}) multiply the two defect operators relevant for the discussion below, which are marginal in the large $N$ limit.

Our main goal is to understand the qualitative structure of the large $N$ conformal manifold of extraordinary surfaces near the origin in Fig.~\ref{fig:eight}. Accordingly, we focus on the large $N$ scaling behavior of the beta functions, rather than their precise numerical coefficients. For this purpose, it suffices to carry out conformal perturbation theory based on \eqref{actionCPTorigin}, and to use the OPE to track the powers of $1/N$ that enter the renormalization of the couplings.

Note that we use the normalization in \eqref{actionCPTorigin} which is different from that in the main text, which will make the $N$-counting more transparent. In this normalization \cite{Goykhman:2020tsk}
\begin{gather}
\label{G_sigmalargeN}
   \langle \sigma(x_1)\sigma(x_2)\rangle
    =\frac{\mathcal{N}_{\sigma}}{|x_{12}|^{2}}\,, \qquad
    \mathcal{N}_{\sigma}=\frac{8}{\pi^2}\,.
\end{gather}
We will also use the following three-point functions \cite{Goykhman:2020tsk}:
\ie
&\langle\sigma^2(x_1)\sigma^2(x_2)\sigma^2(x_3)\rangle=\frac{8\mathcal{N}_{\sigma}^{\,3}}{\left(|x_{12}||x_{13}||x_{23}|\right)^{\Delta_{\sigma^2}}}\,, 
\quad \langle \sigma^2(x_1)\sigma(x_2)\sigma(x_3)\rangle=\frac{2\mathcal{N}_{\sigma}^{\,2}}{|x_{12}|^{\Delta_{\sigma^2}}|x_{13}|^{\Delta_{\sigma^2}}|x_{23}|^{2\Delta_{\sigma}-\Delta_{\sigma^2}}}\,.\label{sss}
\fe 
Here we keep only the leading-order normalization factors in $N$.

To determine which structures can renormalize these couplings, it is enough to study one-point functions of $\sigma^2(z,\bold{0})$ and $\sigma(z,\bold{0})$ in the presence of (and away from) the defect insertions. Indeed, logarithmic divergences in these one-point functions correspond to local counterterms proportional to $\hat{\sigma}^2$ and $\partial_z \hat{\sigma}$ on the defect. We therefore consider connected correlators of the form
\begin{gather}
    \Bigl\langle
    \sigma^2(z,\bold{0})\,
    \prod_{a=1}^{2k}\partial_z \hat{\sigma}(\bold{x}_a)\,
    \prod_{b=1}^{M}\hat{\sigma}^2(\bold{y}_b)
    \Bigr\rangle_{\rm conn}\,,
    \quad 
    \Bigl\langle
    \sigma(z,\bold{0})\,
    \prod_{a=1}^{2k-1}\partial_z \hat{\sigma}(\bold{x}_a)\,
    \prod_{b=1}^{M}\hat{\sigma}^2(\bold{y}_b)
    \Bigr\rangle_{\rm conn}\,.
    \label{sigmacorrapp}
\end{gather}
This is because only the connected parts are relevant for the renormalization of $\hat{h}_0$ and $\hat{h}_{\rm m,0}$. The disconnected pieces arise from the identity operator in intermediate OPE channels. Thus, throughout this appendix we subtract the identity contributions.

The large-$N$ counting can be deduced from the following schematic OPE rules. Since we only care about the powers of $N$, we suppress the position dependence, which are uniquely fixed by conformal symmetry. 
For large $N$ power counting, it is enough to retain the least suppressed non identity channels (assuming no accidental cancellations), 
\ie 
    \hat{\sigma}^2 \times \hat{\sigma}^2
    \sim
    1
    +\,\hat{\sigma}^2
    + \cdots\,,
   ~~
    \partial_z \hat{\sigma} \times \partial_z \hat{\sigma}
    \sim
    1
    + \frac{\hat{\sigma}^2}{N}
    + \cdots\,,
  ~~
    \sigma \times \partial_z \hat{\sigma}
    \sim
    1
    + \frac{\hat{\sigma}^2}{N}
    + \cdots\,.
    \label{operule}
\fe 
The first relation follows from the three-point function $\langle \sigma^2 \sigma^2 \sigma^2\rangle$ in \eqref{sss}, while the second and third follow from $\langle \sigma^2 \sigma \sigma\rangle$ in \eqref{sss} after differentiation and bringing the bulk points to the defect. Equivalently, for the purpose of large-$N$ power counting one may use the replacement rules in the correlation functions \eqref{sigmacorrapp},
\begin{gather}
    \hat{\sigma}^2 \hat{\sigma}^2 \;\rightarrow\; \hat{\sigma}^2\,,
    \qquad
    \partial_z \hat{\sigma}\,\partial_z \hat{\sigma}
    \;\rightarrow\;
    \frac{\hat{\sigma}^2}{N}\,,
    \qquad
    \sigma\,\partial_z \hat{\sigma}
    \;\rightarrow\;
    \frac{\hat{\sigma}^2}{N}\,,
    \label{replacementrulesapp}
\end{gather}
where as before we suppress the position dependence and only retain the operator content and the $N$ scaling.

The second basic ingredient is the standard large-$N$ scaling of connected correlators of the normalized singlet operator $\sigma^2$:
\begin{gather}
    \bigl\langle \sigma^2(x_1)\cdots \sigma^2(x_L)\bigr\rangle_{\rm conn}
    \sim
    N^0\,.
    \label{sigmasqscalingapp}
\end{gather}
Combining \eqref{replacementrulesapp} with \eqref{sigmasqscalingapp}, we can determine the large-$N$ order of the correlators in \eqref{sigmacorrapp}.

Let us first consider the renormalization of $\hat{h}_0$. A contribution with $2k$ insertions of $\partial_z \hat{\sigma}$ and $M$ insertions of $\hat{\sigma}^2$ is weighted by $\hat{h}_{\rm m,0}^{2k}\hat{h}_0^M$.  Using \eqref{replacementrulesapp}, one channel which may contribute can be realized as follows. First, the $2k$ factors of $\partial_z \hat{\sigma}$ can be reduced pairwise to $k$ insertions of $\hat{\sigma}^2$, and each such reduction gives one power of $1/N$. After these reductions, the correlator is a connected correlator of $\hat{\sigma}^2$ and is therefore of order $N^0$ by \eqref{sigmasqscalingapp}. Hence 
\begin{gather}
    \hat{h}_{\rm m,0}^{2k} \hat{h}_0^M
    \int
    \prod_{a=1}^{2k} d^2 \bold{x}_a
    \prod_{b=1}^{M} d^2 \bold{y}_b\,
    \Bigl\langle
    \sigma^2(z,\bold{0})\,
    \prod_{a=1}^{2k}\partial_z \hat{\sigma}(\bold{x}_a)\,
    \prod_{b=1}^{M}\hat{\sigma}^2(\bold{y}_b)
    \Bigr\rangle_{\rm conn}
    \sim
    \frac{\hat{h}_{\rm m,0}^{2k} \hat{h}_0^M}{N^k}\,.
    \label{hscalingapp}
\end{gather}
Other channels may also contribute to the beta functions. Nevertheless, after removing the identity contribution, the channel considered above is the least suppressed at large $N$. Therefore no other channel can lead to a lower power of $N$: at most, it can contribute at the same order or with additional suppression in $\tfrac{1}{N}$.

The analysis for $\hat{h}_{\rm m,0}$ is analogous. A contribution with $2k-1$ insertions of $\partial_z \hat{\sigma}$ and $M$ insertions of $\hat{\sigma}^2$ is weighted by $\hat{h}_{\rm m,0}^{2k-1}\hat{h}_0^M$. In this case, we first use the mixed OPE in \eqref{operule} to combine the external bulk field $\sigma(z,\bold{0})$ with one insertion of $\partial_z \hat{\sigma}$. This produces one factor of $1/N$ and replaces the pair by $\hat{\sigma}^2$. The remaining $2k-2$ insertions of $\partial_z \hat{\sigma}$ are then reduced pairwise using \eqref{operule}, yielding an additional factor of $1/N^{k-1}$. The resulting connected correlator is again of order $N^0$. Hence
\begin{gather}
    \hat{h}_{\rm m,0}^{2k-1} \hat{h}_0^M
    \int
    \prod_{a=1}^{2k-1} d^2 \bold{x}_a
    \prod_{b=1}^{M} d^2 \bold{y}_b\,
    \Bigl\langle
    \sigma(z,\bold{0})\,
    \prod_{a=1}^{2k-1}\partial_z \hat{\sigma}(\bold{x}_a)\,
    \prod_{b=1}^{M}\hat{\sigma}^2(\bold{y}_b)
    \Bigr\rangle_{\rm conn}
    \sim
    \frac{\hat{h}_{\rm m,0}^{2k-1} \hat{h}_0^M}{N^k}\,.
    \label{cscalingapp}
\end{gather}

Consequently the most general form of the beta functions consistent with large-$N$ counting is, at leading order in $N$,
\begin{gather}
    \beta_{\hat{h}}
    =
    \gamma_{\hat{\sigma}^2}\, \hat{h}
    +
    \sum_{M\geq 2} a_{0,M}\, \hat{h}^M
    +
    \sum_{k\geq 1}\sum_{M\geq 0}
    a_{k,M}\,
    \frac{\hat{h}_{\rm m}^{2k} \hat{h}^M}{N^k}\,,
~~
    \beta_{\hat{h}_{\rm m}}
    =
    \gamma_{\partial_z \hat{\sigma}}\, \hat{h}_{\rm m}
    +
    \sum_{k\geq 1}\sum_{M\geq 0}
    b_{k,M}\,
    \frac{\hat{h}_{\rm m}^{2k-1} \hat{h}^M}{N^k}\,,
    \label{betachgeneralapp}
\end{gather}
with $b_{1,0}=0$. 
Here $\gamma_{\hat{\sigma}^2} \sim \tfrac{1}{N}$ and $\gamma_{\partial_z \hat{\sigma}}\sim \tfrac{1}{N}$  denote the bulk anomalous dimensions of the corresponding operators, while the coefficients $a_{k,M}$ and $b_{k,M}$ are numerical constants that require an explicit computation of logarithmic divergences. We caution that some of the coefficients $a_{k,M},b_{k,M}$ may be zero due to subtle cancellations. Indeed, one finds that $a_{0,M}=0$ for $M>2$ to leading order in $N$ \cite{Ord}.

The beta functions \eqref{betachgeneralapp} suggest two qualitatively different possibilities for defect fixed points with different $N$ scaling. One possibility is a perturbative branch with
\begin{gather}
    \hat{h}_* \sim \frac{1}{N}\,,
    \qquad
    c_* = 0\,,
\end{gather}
for which the expansion is organized in powers of $1/N$. This describes the ordinary surface defect at large $N$ which we study in detail in \cite{Ord}. A second possibility is
\begin{gather}
    \hat{h}_* \sim N^0\,,
    \qquad
    \hat{h}_{\rm m,\star} \sim \sqrt{N}\,.
\end{gather}
 In this regime, all terms in beta functions \eqref{betachgeneralapp} become parametrically comparable, and the perturbative large-$N$ expansion on the defect is no longer controlled. This is the regime which we study in Section~\ref{sec:largeN} by directly solving the conformal defects non-perturbatively in the defect couplings at large $N$.

Next we will connect the perturbative analysis here with the non-perturbative solutions in Section~\ref{sec:largeN} (summarized in Fig.~\ref{fig:eight}) by expanding around the origin in the coupling space, namely in the regime $\hat{h}\ll 1$ and $\tfrac{\hat{h}_{\rm m}}{\sqrt{N}}\ll 1$ at the leading order in large $N$, to explain the ``cross'' in Fig.~\ref{fig:eight}. In this limit the anomalous dimensions in \eqref{betachgeneralapp} can be neglected, since they are of order $\tfrac{1}{N}$, and we may also drop higher-order terms in $\hat{h}$ and $\tfrac{\hat{h}_{\rm m}}{\sqrt{N}}$. The beta functions up to subleading terms are
\begin{align}
&\beta_{\hat{h}}=a_{0,2}\hat{h}^2+a_{1,0}\frac{\hat{h}_{\rm m}^2}{N}+\dots\,,\quad \beta_{\hat{h}_{\rm m}}=0 +\dots \,.
       \label{betaCross}
\end{align}
This immediately explains the branch structure at the origin in Fig.~\ref{fig:eight}: to leading order, there are two lines of fixed points parametrized by $\hat{h}_{\rm m}=\hat{h}_{\rm m,\star}$, for each possible values for $\hat{h}$ at
\begin{gather}
    \hat{h}_\star = \pm \sqrt{-\frac{a_{1,0}}{a_{0,2}}}\,\frac{|\hat{h}_{\rm m,\star}|}{\sqrt{N}} \,,
    \label{RelBetweenHandC}
\end{gather}
which join at the origin at $\hat{h}_\star=\hat{h}_{\rm m,\star}=0$. We will fix the ratio of the beta function coefficients in \eqref{RelBetweenHandC} below which may be checked by explicit conformal perturbation theory.

To determine on which side of the $h=0$ axis the branches $D_\mu^{\rm I}$ and $D_\mu^{\rm II}$ lie near the origin in Fig.~\ref{fig:eight}, we must relate the couplings $\hat{h}$ and $\hat{h}_{\rm m}$ to $\mu$. In what follows we work at leading order in the couplings, so we may use renormalized couplings throughout. We begin with the relation between $\hat{h}_{\rm m}$ and $\mu$. Since we know that $\langle \sigma(z,\bold{0})\rangle=\frac{\sqrt{N}\mu}{z}$, the leading contribution to this one-point function gives
\begin{gather}
    \langle  \sigma(z,\bold{0})\rangle=\hat{h}_{\rm m}\int d^2x_2\langle \sigma(z,\bold{0})\partial_{z_2} \sigma(z_2,\bold{x}_2) \rangle\big|_{z_2=0}=\frac{2\pi\mathcal{N}_{\sigma}}{z}\hat{h}_{\rm m}\,,
\end{gather}
from which it follows that
\begin{gather}
    \frac{\hat{h}_{\rm m,\star}}{\sqrt{N}}=\frac{\mu}{2\pi  \mathcal{N}_\sigma}\,.
    \label{CinTermsOfmu}
\end{gather}
Next, we determine the anomalous dimension of $\hat{\sigma}$. The leading contribution comes from
\begin{gather}
    \hat{h} \int d^2 \bold{x}_3\langle \hat \sigma(\bold{x}_1)\hat \sigma(\bold{x}_2)\hat{\sigma}^2(\bold{x}_3)\rangle=2\mathcal{N}^2_\sigma \hat{h}\int d^2\bold{x}_3\frac{1}{(\bold{x}_3-\bold{x}_1)^2(\bold{x}_3-\bold{x}_2)^2}=8 \pi \mathcal{N}^2_{\sigma}\hat{h}\frac{\log{\Lambda|\bold{x}_1-\bold{x}_2|}}{(\bold{x}_1-\bold{x}_2)^2}\,,
\end{gather}
where we have used the three-point function \eqref{sss} and the results of Appendix C in \cite{Krishnan:2023cff}. This implies
\begin{gather}
    \langle \hat \sigma(\bold{x}_1)\hat \sigma(\bold{x}_2)\rangle=\frac{\mathcal{N}_\sigma}{(\bold{x}_1-\bold{x}_2)^2}
\left(1+8\pi \mathcal{N}_\sigma \hat{h}\log{\Lambda|\bold{x}_1-\bold{x}_2|}\right)\,,
\end{gather}
from which we read off the anomalous dimension
\begin{gather}
    \gamma_{\hat{\sigma}}=-4\pi \mathcal{N}_\sigma \hat{h}\,.
    \label{AnomSigmaHatMu}
\end{gather}
This relation immediately fixes the position of $D^{\rm I}_{\mu}$ and $D^{\rm II}_{\mu}$ in  Fig.~\ref{fig:eight}. Finally, combining \eqref{RelBetweenHandC} with \eqref{CinTermsOfmu}, we obtain
\begin{gather}
    \gamma_{\hat{\sigma}}=\mp 2|\mu|\sqrt{-\frac{a_{1,0}}{a_{0,2}}} \,.
\end{gather}
On the other hand, in section \ref{sec:largeN}, we find a parity odd scalar operator on the defect with scaling dimension $1+2\mu$ or $1-2\mu$, depending on the branch of fixed points, $D^{\rm I}_{\mu}$ or $D^{\rm II}_{\mu}$. It is natural to identify this operator with $\hat\sigma$ in the vicinity of the $h=h_{\rm m}=0$.
This identification leads to 
\begin{gather}
    a_{1,0}=-a_{0,2}=4\pi \mathcal{N}_{\sigma}\,,
\end{gather}
where the second equality is determined by standard conformal perturbation theory in \cite{Ord}.

\section{Normal Surface,  Defect Fusion and Large $N$}\label{app:norm}
In the main text, we have focused on the extraordinary surface defined by the gradient deformation to the GNY Lagrangian $L_{\rm ext} = h_{\rm m} \pa_z\phi\delta(z)$. Here we consider the case with a deformation by the pseudoscalar itself
$L_{\rm norm} = h_{\rm o} \phi\delta(z)$ which defines the normal surface defect, which breaks all the discrete parity symmetries.  Again, from general consideration based on the IR factorization, anomalies and $b$-theorem, as discussed in detail in Section~\ref{sec:anomaly}, the solutions to this defect RG flow are fixed to be
\ie
\begin{array}{@{}r@{\hspace{0.8em}}c@{\hspace{1em}}l@{}}
\multirow[c]{2}{*}[0ex]{$\cD_{\rm norm} \equiv \bigl[e^{h_{\rm o}\int_{z=0}\phi}\bigr]_{\rm ren}$}
& \raisebox{0ex}{\smash{\rotatebox[origin=c]{10}{$\xrightarrow{\;h_{\rm o}>0\;}$}}}
& |B_1\rangle\langle \overline{B_1}|  \\[1ex]
& \raisebox{0ex}{\smash{\rotatebox[origin=c]{-10}{$\xrightarrow{\;h_{\rm o}<0\;}$}}}
& |\overline{B_1}\rangle\langle B_1| 
\end{array}
\label{normsol}
\fe
In particular, in the absence of emergent $2d$ chiral fermions, the left boundary $|B_1\ra$ must be paired with $\la \overline{B_1}|$ instead of $\la B_1|$ in order to cancel the defect gravitational anomaly \cite{Wang:2020xkc}.

One can also infer \eqref{normsol} by relating to RG flow on a strip similar to that in Fig.~\ref{fig:squeeze} with a constant mass profile. The constant $\phi$ deformation in the strip breaks parity symmetry explicitly and produces a non-degenerate gapped ground state which leads to the simple factorization channels in \eqref{normsol}.

Furthermore, there are curious relations between the extraordinary surface defect and the normal surface defect via defect fusion  \cite{Bachas:2007td, Bachas:2013ora, Konechny:2015qla,Soderberg:2021kne,  SoderbergRousu:2023zyj,Diatlyk:2024zkk,Kravchuk:2024qoh, Cuomo:2024psk}: focusing on the conformal normal and extraordinary surfaces defined by $h_{\rm o},h_{\rm m}>0$ in the UV, we have the following fusion algebra (which are not associative \cite{Diatlyk:2024zkk}),
\ie \label{fusion}
&\cD_{\rm norm} \circ \overline{\cD}_{\rm norm} = \cD_{\rm ext}\,,\quad 
\overline{\cD}_{\rm norm} \circ  {\cD}_{\rm norm} = \overline{\cD}_{\rm ext}\,,
\quad 
\overline{\cD}_{\rm ext}\circ {\cD}_{\rm ext}=\overline{\cD}_{\rm norm}\,,\quad {\cD}_{\rm ext}\circ \overline{\cD}_{\rm ext}=\cD_{\rm norm}
\,,\quad 
\\
&\cD_{\rm norm} \circ \cD_{\rm ext} = \cD_{\rm ext} \circ \overline{\cD}_{\rm norm} =\cD_{\rm ext}\,,\quad 
\overline{\cD}_{\rm norm} \circ \overline{\cD}_{\rm ext} = \overline{\cD}_{\rm ext} \circ  {\cD}_{\rm norm} =\overline{\cD}_{\rm ext} \,,\quad 
\\
&
\overline{\cD}_{\rm norm} \circ \cD_{\rm ext} = \overline{\cD}_{\rm norm}\,,\quad \cD_{\rm ext} \circ {\cD}_{\rm norm} =  {\cD}_{\rm norm}\,, 
\fe
where $\overline{\cD}$ denotes the time-reversal of the defect $\cD$ and all defects above are fusion idempotents which is a general property of generalized pinning defects \cite{Popov:2025cha}. 

The fusion rules above follow straightforwardly from the IR descriptions of the corresponding surfaces in terms of factorized boundaries in \eqref{mainsol} and \eqref{normsol}, by analyzing the low energy spectrum between a pair of adjacent boundaries
as discussed in \cite{Diatlyk:2024zkk}. When the adjacent boundaries for a sandwich do not align, such as in $\la B_1 | \overline{B_1}\ra$, we expect a unique gapped ground state, and this explains the first fusion product in the second line of \eqref{fusion}. On the other hand, when the boundaries align, such as in $\la B_1|B_1\ra$, we expect gapless modes described by $N$ anti-chiral Majorana fermions on the surface, as is required by the defect gravitational anomaly (see towards the end of Section~\ref{sec:anomaly}). Similarly, the fusion $\la \overline{B_1}|\overline{B_1}\ra$ produces $N$  chiral Majorana surface fermions. They explain the first two fusion products in the first line of \eqref{fusion}. Taking into account that chiral and anti-chiral surface fermions can be lifted together by ${\rm O}(N)$ and parity symmetric surface mass terms, one can deduce the rest of the fusion products in \eqref{fusion}.

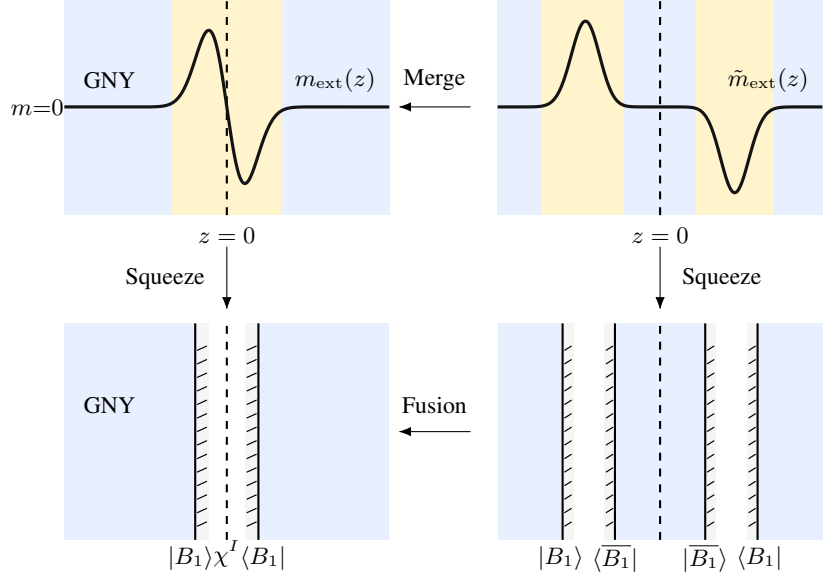
\begin{figure}[!htb]
\centering
{
\begin{tikzpicture}[
    x=1.2cm,y=1.2cm,
    >=Latex,
    every node/.style={font=\small},
    panel/.style={line width=1.05pt, line cap=round, line join=round},
    guide/.style={line width=0.8pt},
    flow/.style={-{Latex[length=3mm,width=2mm]}, line width=0.95pt},
    hatch/.style={line width=0.45pt}
]

\definecolor{gnyfill}{RGB}{232,240,255}
\definecolor{stripfill}{RGB}{255,244,204}
\definecolor{shadefill}{RGB}{245,245,245}
\definecolor{massline}{RGB}{20,20,20}

\begin{scope}[shift={(0,3.6)}]
    \fill[gnyfill]   (0,0) rectangle (1.20,2.4);
    \fill[stripfill] (1.20,0) rectangle (2.40,2.4);
    \fill[gnyfill]   (2.40,0) rectangle (3.60,2.4);

    \draw[guide,dashed] (1.80,0)--(1.80,2.4);

    \draw[line width=1.2pt, color=massline, samples=300, domain=0:3.6]
        plot (\x,{1.2 - 7*(\x-1.8)*exp(-12.5*(\x-1.8)^2)});
  \node at (.5,1.5) {GNY};
    \node at (3,1.5) {$m_{\rm ext}(z)$};
    \node at (1.80,-0.22) {$z=0$};
    \node at (-.3,1.2) {$m{=}0$};
\end{scope}

\begin{scope}[shift={(4.8,3.6)}]
    \fill[gnyfill]   (0,0) rectangle (0.50,2.4);
    \fill[stripfill] (0.5,0) rectangle (1.4,2.4);
    \fill[gnyfill]   (1.4,0) rectangle (2.2,2.4);
    \fill[stripfill] (2.2,0) rectangle (3.1,2.4);
    \fill[gnyfill]   (3.06,0) rectangle (3.60,2.4);

    \draw[guide,dashed] (1.80,0)--(1.80,2.4);

    \draw[line width=1.2pt, color=massline, samples=400, domain=0:3.6]
        plot (\x,{
            1.2
            + 0.98*exp(-((\x-0.98)^2)/(2*0.16^2))*(1-exp(-((\x-1.8)^2)/(0.44^2)))
            - 0.98*exp(-((\x-2.62)^2)/(2*0.16^2))*(1-exp(-((\x-1.8)^2)/(0.44^2)))
        });

    \node at (3.,1.5) {$\tilde m_{\rm ext}(z)$};
    \node at (1.80,-0.22) {$z=0$};
\end{scope}

\begin{scope}[shift={(0,0)}]
    \fill[gnyfill]   (0,0) rectangle (1.45,2.4);
    \fill[shadefill] (1.45,0) rectangle (1.60,2.4);
    \fill[shadefill] (2.00,0) rectangle (2.15,2.4);
    \fill[gnyfill]   (2.15,0) rectangle (3.60,2.4);

    \draw[guide] (1.45,0)--(1.45,2.4);
    \draw[guide] (2.15,0)--(2.15,2.4);

    \foreach \y in {0.15,0.30,...,2.20}{
        \draw[hatch] (1.47,\y)--(1.58,\y+0.05);
        \draw[hatch] (2.02,\y)--(2.13,\y+0.05);
    }

    \draw[guide,dashed] (1.80,0)--(1.80,2.4);

    \node at (1.8,-.15) { $\chi^I$};
    \node at (2.2,-.2) { $\langle B_1|$};
    \node at (1.4,-.2) { $|B_1\rangle$};

      \node at (.5,1.5) {GNY};
\end{scope}

\begin{scope}[shift={(4.8,0)}]
    \fill[gnyfill]   (0.00,0) rectangle (0.72,2.4);
    \fill[shadefill] (0.72,0) rectangle (0.84,2.4);
    \fill[shadefill] (1.18,0) rectangle (1.30,2.4);
    \fill[gnyfill]   (1.30,0) rectangle (2.30,2.4);
    \fill[shadefill] (2.30,0) rectangle (2.42,2.4);
    \fill[shadefill] (2.76,0) rectangle (2.88,2.4);
    \fill[gnyfill]   (2.88,0) rectangle (3.60,2.4);

    \draw[guide] (0.72,0)--(0.72,2.4);
    \draw[guide] (1.30,0)--(1.30,2.4);
    \draw[guide] (2.30,0)--(2.30,2.4);
    \draw[guide] (2.88,0)--(2.88,2.4);

    \foreach \y in {0.15,0.30,...,2.20}{
        \draw[hatch] (0.74,\y)--(0.82,\y+0.05);
        \draw[hatch] (1.20,\y)--(1.28,\y+0.05);
        \draw[hatch] (2.32,\y)--(2.40,\y+0.05);
        \draw[hatch] (2.78,\y)--(2.86,\y+0.05);
    }

    \draw[guide,dashed] (1.80,0)--(1.80,2.4);

    \node at (1.3,-0.20) {$\langle \overline{B_1}|$};
    \node at (.7,-0.20) {$|B_1\rangle$};

    \node at (2.9,-0.20) {$\langle {B_1}|$};
    \node at (2.3,-0.20) {$|\overline{B_1}\rangle$};
\end{scope}

\draw[->] (4.5,4.80) -- (3.7,4.80);
\node at (4.10,5.10) {Merge};

\draw[->] (4.5,1.20) -- (3.7,1.20);
\node at (4.10,1.50) {Fusion};

\draw[->] (1.80,3.25) -- (1.80,2.55);
\node[left] at (1.65,2.90) {Squeeze};

\draw[->] (6.60,3.25) -- (6.60,2.55);
\node[right] at (6.75,2.90) {Squeeze};

\end{tikzpicture}
}
\caption{Fusion of normal surface defects in the GNY CFT into the extraordinary surface defect, shown in the bottom diagrams. This fusion rule is reflected in the corresponding fermion mass profiles that engineer these defect configurations, shown in the top diagrams. The top and bottom diagrams are related by squeezing the configuration in the yellow regions.}
\label{fig:fusion}
\end{figure}

Conversely, without directly invoking the IR solutions \eqref{mainsol} and \eqref{normsol}, one can argue for these fusion rules by relating them to bulk deformations and taking appropriate limits, as illustrated in Fig.~\ref{fig:fusion}. In particular, this provides an alternative route to infer the IR solution for the extraordinary surface \eqref{mainsol} from that of the normal surface \eqref{normsol}, given that the former is constrained to arise as the fusion product of the latter with its orientation reversal in the IR.

To see this, we proceed in several steps. First, the composite configuration of the normal defect $\cD_{\rm norm}$ and its orientation reversal $\overline{\cD}_{\rm norm}$ can be obtained by squeezing the fermion mass profile $\tilde m_{\rm ext}$ shown in the top right diagram of Fig.~\ref{fig:fusion}. In this squeezing limit, the localized bumps generate effective couplings to the pseudoscalar $\phi$ at fixed locations along the $z$ direction. 
Second, up to deformations that are irrelevant at long distances, the mass profile $m(z)$ in Fig.~\ref{fig:squeeze}, which defines the extraordinary surface upon squeezing, may be replaced by $m_{\rm ext}(z)$ in the top right diagram of Fig.~\ref{fig:fusion}. 
Third, the profile $\tilde m_{\rm ext}$ defining the composite defect is related to $m_{\rm ext}$ by smoothly opening a plateau at $m=0$ near the core of the domain wall at $z=0$. As a result, for scaling observables at large transverse separations (from the bumps, and equivalently to the defect locations), these profiles are indistinguishable.

Taken together, this implies that the IR limit from the top right diagram to the bottom left diagram can be taken in two equivalent ways. By first merging and then squeezing the bumps in $\tilde m_{\rm ext}(z)$, one recovers the first fusion rule in \eqref{fusion}. This provides a nontrivial consistency check between \eqref{mainsol} and \eqref{normsol}.

Finally we can repeat the large $N$ analysis performed in Section~\ref{sec:largeN} in the case of the normal surface defect and directly confirm \eqref{normsol}. The key difference is that the bulk one-point function of $\phi$, taking the following at large $N$
form 
\begin{align}
  \langle\phi (z, \bold{x})\rangle = \frac{\mu}{|z|}\,,
\end{align}
compared to \eqref{sigma1pt}, reflecting the breaking of transverse parity symmetry by the normal surface defect. Performing the same folding trick yields $N$ fermions $\Psi_1^I$ of mass $\mu$ and $N$ fermions $\Psi_2^I$ of mass $-\mu$ in $\mathbb{H}_3$. The difference in mass immediately implies that the off-diagonal two-point functions $G_{12}$ vanish. The diagonal two-point $G_{11}$ is the same as in \eqref{12pm} because it is completely fixed by the bulk data. $G_{22}$ is related to $G_{11}$ by flipping the sign of $\mu$. Below, we argue that instead of a line of fixed points, we only have the trivial defect $\mu=0$ and an additional fixed point at $\mu=\frac{1}{2}$. For $0<\mu<\frac{1}{2}$, using the general O$(N)$ fermion boundary condition \eqref{generalbc} and $G_{12}\equiv 0$, we find 
\begin{align}\label{bc1}
    \gamma^1 G_{11} = \sin(\vartheta)\, G_{11}= -\frac{\sin(\vartheta)}{16\pi}\left(G_++G_-\right)\,, \quad z_1\to 0\,,
\end{align}
where $z_1$ is the transverse coordinate of the first point. On the other hand, the properties of $G_\pm$ given by \eqref{gammaG} yield \begin{align}\label{bc2}
    \gamma^1 G_{11} =  -\frac{1}{16\pi}\left(-G_++G_-\right)\,, \quad z_1\to 0\,.
\end{align}
Since $G_\pm$ are linearly independent solutions for $0<\mu<\frac{1}{2}$, there does not exist an angle $\vartheta$ for \eqref{bc1} and \eqref{bc2} to be compatible. This argument does not work for $\mu=\frac{1}{2}$ because $G_\pm$ coincide at this point. At $\mu=\frac{1}{2}$, $G_{11}\propto G_+$, implying $\vartheta=0$ in \eqref{generalbc}. We can perform exactly the same analysis for $\Psi_2^I$. Altogether, the positive $\phi$ deformation  (i.e. $h_{\rm o}>0$) on the interface leads to the $|B_1\rangle\langle \overline{B_1}|$ fixed point, described by fermions $\Psi_1^I$ and $\Psi_2^I$ of opposite masses $\pm\frac{1}{2}$ with the boundary condition
\begin{align}
    \left.\gamma^1 \Psi_1^I \right|_{z= 0} = - \left.\Psi_1^I\right|_{z= 0}, \quad \left.\gamma^1 \Psi_2^I \right|_{z=0} =  \left.\Psi_2^I\right|_{z=0}~.
\end{align}
Similarly, the negative $\phi$ deformation leads to the fixed point $|\overline{B_1}\rangle\langle B_1|$ related by a transverse reflection.
These flows are obviously consistent with the $b$-theorem.

\section{Harmonic analysis in $\mathbb{H}_{3}$}
\label{AdSreview}

The three-dimensional hyperbolic is a hypersurface in the ambient space $\mathbb R^{1,3}$
\begin{align}
    \eta_{AB} X^A X^B = -L^2\,,
\end{align}
where $\eta_{AB} = \text{diag}(-,+,+,+)$ and $L$ is the radius. We will take the radius to be 1.  $\mathbb{H}_3$ in the Poincar\'e coordinates is related to the upper half plane in flat space by a Weyl transformation:
\begin{align}\label{Pcoord}
    X^0 =\frac{z^2+\bold{x}^2+1}{2z}\,,\quad X^1 = \frac{x^1}{z}\,, \quad X^2 = \frac{x^2}{z}\,, \quad X^3 = \frac{z^2+\bold{x}^2-1}{2z}\,,
\end{align}
with the metric being $ds^2 =\frac{dz^2+d\bold{x}^2}{z^2}$. The boundary is at $z=0$.
Another useful coordinate system is the the global coordinate:
\begin{align}\label{Gcoord}
    X^0=\cosh(r)\,, \quad X^a =\sinh(r) \omega^a\,, \quad \omega^a \in S^2\,,
\end{align}
and the corresponding metric reads $ds^2=dr^2 + \sinh^2(r) d\omega^2$. So the boundary in this coordinate system is a two-sphere, whose radius can be used as an IR regulator of the volume. Let $R$ be the radius of the boundary sphere and then the {\it regularized} volume of $\mathbb{H}_3$ becomes \cite{Giombi:2013yva,Casini:2010kt,Casini:2011kv,Diaz:2007an}
\begin{align}\label{AdSvol}
V_3 = -2\pi \log R\,.
\end{align}

The Laplacian operator of  $\mathbb{H}_3$ has a continuous spectrum $[1, \infty)$ \cite{Camporesi:1994ga}. We denote the eigenmodes by $\psi_{\nu\ell}(X)$:
\begin{align}
    -\nabla^2 \psi_{\nu\ell} = (1+\nu^2)\psi_{\nu\ell}\,,
\end{align}
where $\nu\ge 0$ and $\ell$ labels angular momenta on $S^2$. The magnetic quantum number is suppressed. We  adopt the following $\delta$-function normalization 
\begin{align}
    \int_X \, \psi^*_{\nu\ell}(X)\psi_{\nu'\ell'}(X) = \delta(\nu-\nu')\delta_{\ell \ell'}\,,
\end{align}
where $\int_X$ is a short-hand notation for the integration over the full $\mathbb H_3$.
Following this normalization, the $\ell=0$ modes take a very simple form 
\begin{align}
    \psi_{\nu 0}(r, \omega) = \frac{1}{\sqrt{2}\pi}\frac{\sin(\nu\, r)}{\sinh(r)}\,.
\end{align}

\subsection{Spectrum decomposition of two-point functions}
Let $G(X_1, X_2)$ be an $\SO(1,3)$ invariant two-point function in $\mathbb{H}_3$. It is a function of $v\equiv -X_1\cdot X_2\ge 1$. We will also use the parameterization $v=\cosh(r)$, where $r$ is the radial coordinate of $X_2$ if we move $X_1$ to $(1,0,0,0)$ by using the $\SO(1,3)$ symmetry.
 For example, the propagator of a free scalar with boundary dimension $\hat\Delta = 1+s$ is $G_s(r)= \frac{e^{-s r}}{4\pi \sinh(r)} $.  $G(X_1, X_2)$ is diagonal in the basis $\psi_{\nu\ell}(X)$ 
\begin{align}
   G(X_1, X_2) = \int_0^\infty d\nu \sum_\ell \rho_G(\nu, \ell) \psi^*_{\nu\ell}(X_1)\psi_{\nu\ell}(X_2)\,.
\end{align}
Using the $\SO(1, 3)$ symmetry, we can place $X_2$ at the origin of the global coordinates and then only the $\ell=0$ term survives because $\psi_{\nu\ell}(r, \omega)\propto \sinh^\ell(r)$. Altogether, the spectral decomposition of $G(X_1, X_2) $ reads 
\begin{align}
    G(X_1, X_2) = \int_0^\infty d\nu\, \rho_G(\nu)\frac{\nu\sin(\nu r)}{2\pi^2\sinh(r)} \,, \quad \rho_G(\nu)\equiv \rho_G(\nu, 0)\,.
\end{align}
It is also conventional to extend the {\it spectral density} $\rho_G(\nu)$ to the whole real line as an even function
\begin{align}\label{specrep}
      G(X_1, X_2) = \int_{-\infty}^\infty d\nu\, \rho_G(\nu) \Omega_\nu(X_1, X_2)\,, \quad \Omega_\nu(X_1, X_2)\equiv \frac{\nu\sin(\nu r)}{4\pi^2\sinh(r)}\,, 
\end{align}
where $\Omega_\nu(X_1, X_2)$ is the (scalar) AdS harmonic functions in three dimensions \cite{Costa:2014kfa}, analogous to Gegenbauer polynomials in sphere. The harmonic functions have the following properties
\begin{align}\label{OC}
    &\text{Orthogonality}:\quad \int_Y \Omega_{\nu}(X, Y)\Omega_{\bar\nu}(Y, Z) = \frac{\delta(\nu-\bar\nu)+\delta(\nu+\bar\nu)}{2}\Omega_\nu(X, Z)\,,\nonumber\\
   & \text{Completeness}:\quad \int_{\mathbb R} d\nu \,\Omega_\nu(X, Y) = \delta(X, Y)\,.
\end{align}
Using these properties, we can invert \eqref{specrep}
\begin{align}\label{invert}
\rho_G(\nu)=\frac{4\pi}{\nu}\int_0^\infty d r \sinh(r)\sin(\nu r)G (r)\,.
\end{align}
Another immediate consequence of \eqref{OC} is that $\rho_{G^{-1}}(\nu) =1/\rho_{G}(\nu)$, where $G^{-1}$ is defined as
    $\int_Y G(X, Y)  G^{-1}(Y, Z) = \delta(X, Z)$\,.

When $G(X, Y)$ is the two-point function of a primary operator  $\CO$ in a BCFT, the poles of $\rho_G(\nu)$ in the lower half $\nu$-plane  determine the boundary scaling dimensions of  operators in the boundary OPE of $\CO$ \cite{Hogervorst:2021uvp,Loparco:2023rug}. Precisely, consider the spectral representation 
$G(X, Y) = \int_{\mathbb R}d\nu\, \rho_G(\nu) \Omega_\nu(X, Y)$, and rewrite the harmonic functions in terms of $G_s$, using the relation
\begin{align}
    \Omega_\nu(X, Y) = \frac{i\nu}{2\pi}(G_{i\nu}(X, Y)-G_{-i\nu}(X, Y))\,.
\end{align}
As we mentioned above, $G_s$ is the propagator of a free scalar of boundary dimension $1+s$. In the meantime, it is also the boundary conformal block corresponding to $\hat\Delta =1+s$. Therefore, we can expand $G(X, Y)$ as follows
\begin{align}\label{bexp}
    G(X, Y)& = \int_{\mathbb R}d\nu\, \rho_G(\nu) \frac{i\nu}{\pi} G_{i\nu}(X, Y)\nonumber\\
    &=\sum_{ \Im(\nu_\star)<0} 2{\rm Res}_{\nu=\nu_\star}(\rho_G)\,\nu_\star G_{i\nu_\star}(X, Y)\,,
\end{align}
where in the last line, we close the contour in the lower-half plane. This expansion indicates boundary operators of dimension $1+i\nu_\star$ for each pole $\nu_\star$ in the lower half-plane.

$\rho_G(\nu)$ can be used to compute the functional determinant of $G(X, Y)$. Following \cite{Camporesi:1994ga}, we consider the heat kernel of $G$
\begin{align}
    K_G(X, Y; \tau) = \int_0^\infty \,d\nu\sum_\ell \psi^*_{\nu\ell}(X)\psi_{\nu\ell}(Y) e^{-\tau\rho_G(\nu, \ell)}\,.
\end{align}
For coincident points, i.e. $X=Y$, the heat kernel is independent of the position $X$ due to the $\SO(1,3)$ symmetry and hence we can move $X$ to the origin of $\mathbb{H}_3$, which leads to the following simplified representation
\begin{align}
    K_G(\tau) \equiv K_G(X,X; \tau)=\frac{1}{2\pi^2 } \int_0^\infty \,d\nu \nu^2 e^{-\tau\rho_G(\nu)}\,.
\end{align}
To regularize the functional determinant, a standard approach is to compute the spectral zeta function
\begin{align}
    \zeta_G(z) = \frac{1}{\Gamma(z)} \int_0^\infty\frac{d\tau }{\tau} \tau^z K_G(\tau) = \frac{1}{2\pi^2 } \int_0^\infty \, \frac{d\nu\, \nu^2}{\rho_G(\nu)^z}\,.
\end{align}
The relation between the functional determinant and  the spectral zeta function is  $  \log \det G = - V_3\zeta_G'(0)$. For instance, given a free scalar of boundary dimension $1+s$ (assuming $s\ge 0$ for simplicity), the spectral density is $\frac{1}{\nu^2+s^2}$ and we find the spectral zeta function $\zeta_s(z) =\frac{ \Gamma \left(-z-\frac{3}{2}\right)}{8 \pi ^{3/2} \Gamma (-z)}s^{2 z+3}$. Its free energy in $\mathbb{H}_3$ is thus 
\begin{align}\label{Fss}
    F_s = \frac{1}{2}V_3 \zeta'_s(0) = \frac{s^3}{6}\log(R)\,,
\end{align}
which can be analytically continued to $s<0$.
Formally dropping the regulator $z$ in the zeta function regularization leads to the following formula of functional determinant in $\mathbb{H}_3$
\begin{align}\label{Fviarho}
   \log \det G = -\frac{\log(R)}{\pi}\int_0^\infty d\nu\, \nu^2 \log \rho_G(\nu)\,.
\end{align}

\subsection{Solving $\langle\delta\phi_a\delta\phi_b\rangle$}\label{app1}
As an application of spectrum decomposition, we solve \eqref{phiphi} for the $\delta\phi$ propagator.
After performing the folding trick and mapping the half-space to hyperbolic space, \eqref{phiphi} translates into 
\begin{equation}\label{KH}
\begin{split}
 &\int d^3x_2 \sqrt{g}\left(K_{11}(x_1, x_2) H_{11}(x_2, x_3)+K_{12}(x_1,x_2)H_{12}(x_2, x_3)\right) =\frac{\delta^3(x_1, x_3)}{\sqrt{g(x_1)}}\\
    &\int d^3x_2 \sqrt{g}\left(K_{11}(x_1, x_2) H_{12}(x_2, x_3)+K_{12}(x_1,x_2)H_{11}(x_2, x_3)\right) = 0
    \end{split}
\end{equation}
where 
\begin{align}
    K_{ab}(x_1, x_2) = \frac{N}{2}\tr\left(G_{ab}(x_1, x_2)G_{ab}(x_2, x_1)\right)\,, \quad  H_{ab}(x_1,x_2) = \langle \delta\phi_a(x_1)\delta\phi_b (x_2)\rangle\,.
\end{align}
Similarly, we define $P$-even and $P$-odd combinations of $\delta\phi_a$
\begin{align}
    \delta\phi_{\rm e} = \frac{\delta\phi_1+\delta\phi_2}{\sqrt{2}}\,, \quad  \delta\phi_{\rm o} = \frac{\delta\phi_1-\delta\phi_2}{\sqrt{2}}\,.
\end{align}
Eq.\eqref{KH} implies that $H_{\rm e}(x_1, x_2)\equiv \langle\delta\phi_{\rm e}(x_1)\delta\phi_{\rm e}(x_2)\rangle$ is the inverse of $K_+\equiv K_{11}+ K_{12}$ and $H_{\rm o}(x_1, x_2)\equiv \langle\delta\phi_{\rm o}(x_1)\delta\phi_{\rm o}(x_2)\rangle$ is the inverse of $K_-\equiv K_{11}- K_{12}$. After a short calculation, we find the explicit form of $K_\pm$ 
\begin{equation}
\begin{split} 
    & \frac{K_+ (x_1, x_2)}{N} =  -\frac{\cosh(2\mu r)\cosh(r)}{16\pi^2\sinh^4(r)}+\frac{\mu\sinh(2\mu r)}{8\pi^2\sinh^3 (r)}\,,\quad  \frac{K_- (x_1, x_2)}{N}=    \frac{1}{32\pi^2}\left(\frac{4\mu^2-1}{\sinh^2(r)}-\frac{2}{\sinh^4(r)}\right)\,.
\end{split}   
\end{equation}
To invert $K_\pm$, we first compute the spectral density of $K_\pm$ using \eqref{invert}. For $K_+$, it reads 
\begin{align}
    \rho_{K_+}(\nu) =  \frac{N}{4\pi\nu}\int_0^\infty d r\left(-\frac{\cosh(2\mu r)\cosh(r)\sin(\nu r)}{\sinh^3(r)}+\frac{2\mu\sinh(2\mu r)\sin(\nu r)}{\sinh^2(r)}\right)\,.
\end{align}
The integral is IR convergent provided $|\mu|<\frac{1}{2}$. We regularize the UV divergence at $r=0$ using analytical continuation. Precisely,  consider 
\begin{align}
\CI_a(\kappa) = \int_0^\infty d r \frac{e^{\kappa r}}{\sinh^a (r)}=\frac{2^{a -1} \Gamma (1-a ) \Gamma \left(\frac{a -\kappa
   }{2}\right)}{\Gamma \left(1-\frac{a+\kappa }{2}\right)}\,,
\end{align}
where the $r$ integral is convergent for $\kappa<a<1$ but the result can be analytically continued in both $\kappa$ and $a$. The first integral in $\rho_{K_+}(\nu)$ is a linear combination of $\CI_a(\pm1\pm 2\mu\pm i\nu)$, with the $a\to 3$ limit being understood. This limit turns out to be finite. The second integral can be treated in the same way. Combining the two integrals, we find
\begin{align}
\rho_{K_+} (\nu) = \frac{N(4 \mu ^2+\nu ^2) }{\cosh
   (\pi  \nu )-\cos (2 \pi  \mu )}\frac{\sinh (\pi  \nu )}{16 \nu}\,.
\end{align}
The spectral density $\rho_{\rm e}$ of $H_{\rm e}$ is equal to $1/\rho_{K_+}$.
 The poles of $\rho_{\rm e}$ encode the boundary operators of $\delta\phi_{\rm e}$. The precise relation between the analytical structure and the boundary channel expansion is given by  \eqref{bexp}. Applying  \eqref{bexp} to $\rho_{\rm e}$ yields 
\begin{align}\label{H+exp}
    &H_{\rm e} (r)=\frac{32}{N\pi}\sum_{n\ge 1}\frac{n^2(1+(-)^{n+1}\cos(2\pi\mu))}{n^2-4\mu^2} G_n(r)\,,
    \end{align}
where $G_s(r)=\frac{e^{-s r}}{4\pi \sinh(r)}$ is the scalar boundary conformal block with 
dimension $\hat\Delta = 1+s$.

Similarly, we derive the spectral density of $K_-$
\begin{align}
\rho_{K_-} (\nu) = \frac{N(4 \mu ^2+\nu ^2) }{\cosh
   (\pi  \nu )+1}\frac{\sinh (\pi  \nu )}{16 \nu}\,.
\end{align}
The spectral density $\rho_{\rm o}$ of $H_{\rm o}$ is given by  $1/\rho_{K_-}$. In contrast to the parity-even case,  $\rho_{\rm o}$ has two $\mu$-dependent poles at $\pm 2i\mu$.
For $0<\mu<\frac{1}{2}$, we can choose either standard or alternate boundary conditions, which yields two inequivalent boundary expansions of $H_{\rm o}$:
    \begin{equation}\label{H-exp}
    \begin{split}
    &H^{\rm I}_{\rm o} (r)=\frac{32\mu\cot(\pi\mu)}{N}G_{2\mu}(r) + \frac{64}{N\pi}\sum_{n\ge 1}\frac{n^2}{ n^2-\mu^2}G_{2n}(r)\,,\\
     &H^{\rm II}_{\rm o} (r)=\frac{32\mu\cot(\pi\mu)}{N}G_{-2\mu}(r) + \frac{64}{N\pi}\sum_{n\ge 1}\frac{n^2}{n^2-\mu^2}G_{2n}(r)\,.
     \end{split}
\end{equation}

\subsection{The $1/N$ correction of free energy}\label{app2}
In the large $N$ expansion, the full free energy of the GN model along the line of  defect fixed point takes the schematic form $F (\mu)= N F_0(\mu)+ F_1(\mu)+F_2(\mu)/N+\cdots$. The leading term corresponds to the one-loop free energy of $\Psi^I_{\rm e/o}$  and vanishes.
 The subleading term is equal to the one-loop free energy of the $\delta\phi_{\rm e/o}$ fields. The zeros of $F'_1(\mu)$ correspond to the defect fixed point for large but finite $N$. Using this method, we can easily recover the defect fixed points of the critical O$(N)$ model found in \cite{Krishnan:2023cff}. 
In the GN case, with the spectral densities calculated above, we compute the free energy $F_1(\mu)$ in AdS using \eqref{Fviarho}. Along the $D^{\rm I}_\mu$ branch, it reads
\begin{align}
    F_{\rm1,  I}(\mu) =  \frac{\log(R)}{2\pi}\int_0^\infty d\nu \,\nu^2 \log\left(\rho_{\rm e}(\nu) \rho_{\rm o }(\nu)\right)~.
\end{align}
Taking derivative respect to $\mu$ yields 
\begin{align}\label{F1d}
   \partial_\mu F_{\rm1,  I}(\mu) &= -\frac{\log(R)}{2\pi}\int_0^\infty d\nu \,\nu^2 \left(\frac{16 \mu }{4 \mu ^2+\nu ^2}+\frac{2 \pi  \sin (2 \pi  \mu )}{\cos (2 \pi  \mu )-\cosh
   (\pi  \nu )}\right)=
   \frac{4}{3}\mu(1+\mu)(1+2\mu)\log(R)\,,
\end{align}
where the 
$\zeta$-function regularization $\frac{1}{\nu^2+4\mu^2}\to \frac{1}{(\nu^2+4\mu^2)^z}$
has been used. Since we work with $\mu\ge 0$, the only critical point of $F_{\rm1,  I}$ is $\mu=0$. On the other hand, along the $D_\mu^{\rm II}$ branch, due to the change of boundary of the leading parity-odd scalar, we have
\begin{align}
   \partial_\mu F_{\rm1,  II}(\mu)
   &=\frac{4}{3}\mu(1-\mu)(1-2\mu)\log(R)\,.
\end{align}
The difference between $\partial_\mu F_{\rm1,  I}$ and $\partial_\mu F_{\rm1,  II}$ at $\mu=\frac{1}{2}$ is related to the loss of analyticity at a crossing point. Nevertheless, given that $\partial_\mu F_{\rm1,  II}$ indeed vanishes at $\mu=0, 1/2$, we believe that 
after taking into account $1/N$ corrections, the fixed lines shrink to two fixed points, namely the defect UV fixed point at $\mu=0$, corresponding to the trivial defect, and the defect IR fixed point at $\mu = \frac{1}{2}$, described by $|B_1\rangle\langle B_1|+\chi^I$.

As a byproduct, we can derive the boundary Weyl anomaly for the $B_1$ phase using \eqref{F1d}. Integrating over $\mu$ from $0$ to $\frac{1}{2}$ in \eqref{F1d} yields $F_{\rm 1, I}(1/2)-F_{\rm 1, I}(0) = \frac{3}{8}\log(R)$. At $\mu=0$, the defect is trivial, so $F_{\rm 1, I}(0)$ vanishes. At $\mu=\frac{1}{2}$, $F_{\rm 1, I}(1/2)$ is twice of the free energy of the $B_1$ boundary, because $\rho_{\rm e}=\rho_{\rm o}$ at $\mu=\frac{1}{2}$ is the spectral density of the $\phi$ fluctuation in the $B_1$ boundary universality class. Therefore, $F_{B_1}^{3d} = \frac{3}{16}\log(R)$, leading to the boundary Weyl anomaly $b_{B_1}^{3d} = -\frac{9}{16}+O(1/N) $.

\section{Anomalous dimension of the chiral fermions near the extraodinary fixed points}\label{ConfPT}
In this appendix, we compute the anomalous dimension of the chiral fermions $\chi^I$ as we turn on the defect coupling $s\int d^2 \bold{x}\,\CO_{\rm e}(\bold{x})$ at the extraordinary fixed point, with $\CO_{\rm e}$ given by \eqref{Oe}.
Here $\psi_L^I, \psi_R^I$ are boundary fermionic operators of dimension $(h, \bar h)= (\frac{1}{2},1)$, belonging to the two decoupled $B_1$ boundary universality classes. The two-point functions of the fermions are normalized as follows
\begin{align}\label{chichi}
    \langle\chi^I(y_1)\chi^J(y_2)\rangle = - \frac{\delta^{IJ}}{y_{12}}\,, \quad \langle\psi_a^I(y_1)\psi_b^J(y_2)\rangle = - \frac{\delta_{ab}\delta^{IJ}}{y_{12}\bar y_{12}^2}\,,
\end{align}
where $y$ denotes the holomorphic coordinate in the $\bold{x}$ plane.

We perform the calculation in the framework of conformal perturbation theory. See, for example \cite{Komargodski:2016auf, Behan:2017emf} for a review on this subject. Due to the vanishing of  $\langle \chi^I\CO_{\rm e}\chi^J\rangle$,
the connected part of the four-point function $\langle \chi^I\CO_{\rm e}\CO_{\rm e}\chi^J\rangle$ gives rise to the leading contribution to the anomalous dimension. Precisely, the anomalous dimension is given by the following regulated integral (no sum over $I$) 
\begin{equation}\label{CPTintegral}
   \begin{split}
    \gamma_\chi &= -\pi s^2 \int_{\mathcal R} d^2 y \bigg(2\langle \chi^I(0)\CO_{\rm e}(y)\CO_{\rm e}(1)\chi^I(\infty)\rangle_c+\langle \chi^I(1)\CO_{\rm e}(0)\CO_{\rm e}(y)\chi^I(\infty)\rangle_c\bigg)\\
    &=-\frac{\pi s^2}{N} \int_{\mathcal R} d^2 y\left(\frac{2}{y(1-\bar y)^2}+\frac{1}{(1-y)\bar y^2}\right)\,,
    \end{split} 
\end{equation}
where $\mathcal R = \{y\in\mathbb C: |y|<1 \,\&\, |y|<|y-1|\}$. The integral of the first term is convergent in the region $\mathcal R$. The second term has a singularity at the origin that can be regularized by the principal value prescription. After evaluating the integrals, we find $
\gamma_\chi = \pi^2 s^2/N$ or equivalent \ie 
(h, \bar h)= \left(\frac{1}{2}+\frac{\pi^2 s^2}{2N},\frac{\pi^2 s^2}{2N}\right)\,.
\label{CPTchiral}
\fe 

Let us also note that the same conformal perturbation theory analysis applies to the anomalous dimension of the operator of dimension $(h,\bar h)=({1\over 2},0)$ in the $\mZ_2$ twisted sector of the Ising$^2$ CFT. In particular, the result \eqref{CPTchiral} can be inferred without performing any integrals, using only the exact solvability of the Ising$^2$ CFT (see \cite{Thorngren:2021yso} for a review of the relevant ingredients).

The Ising$^2$ CFT can be viewed as a bosonization of two Majorana fermions, whose chiral components we denote by $\lambda_a$ and antichiral components by $\tilde\lambda_a$ with $a=1,2$. The energy (mass) operators of the two Ising models are $\ep_a\equiv \lambda_a\tilde\lambda_a$, and the product operator $\ep_1\ep_2$ is exactly marginal. It generates the orbifold branch of the $c=1$ compact boson CFT parametrized by the radius $R$, with the Ising$^2$ point located at $R=2$.
Explicitly, the $c=1$ compact boson has action \cite{Polchinski:1998rq} with $\theta\sim \theta+2\pi$,
\ie
S_{\rm boson}={R^2\over 2\pi }\int d^2 \bold{x}\,\pa \theta \bar\pa \theta\,,
\label{bosact}
\fe
and the charge-conjugation orbifold imposes further that $\theta\sim -\theta$.

From the bosonic perspective, the fermion operator $\lambda_1$ is identified with an operator in the sector twisted by a $\mZ_2$ symmetry preserved along the orbifold branch, which acts as $\theta \to \theta +\pi$. In terms of the charge-conjugation-even momentum-winding operators $V^+_{n,w}$ of the compact boson orbifold (see \cite{Thorngren:2021yso} for notation), one has $\lambda_1=V^+_{1,1/2}$ with conformal dimension
\ie
(h,\bar h)=\left({1\over 2}\left({R\over 4}+{1\over R}\right)^2\,,{1\over 2}\left({R\over 4}-{1\over R}\right)^2\right)\,.\label{bosans}
\fe
Note that the half winding is allowed since this is an operator in the twisted sector of $\pi$-shift in $\theta$ and thus only required to be mutually local with even winding operators.

The problem of determining the anomalous dimension of $\chi^I$ in our surface defect setup at large $N$ is mathematically identical to that of $\lambda_1$ in the Ising$^2$ CFT along the orbifold branch. More explicitly, the operator $\cO_{\rm e}$ is replaced by $\lambda_1(\tilde\lambda_1\ep_2)=\ep_1\ep_2$. It remains to identify the map between the marginal couplings $s$ and $R$.

For fixed $I$, the effective coupling between the canonically normalized marginal operator and $\chi^I$ is ${s\over \sqrt{N}}$ from \eqref{Oe}. On the other hand, the canonically normalized marginal operator in the compact boson orbifold is $R^2 \pa \theta \bar\pa \theta$. Expanding \eqref{bosact} around $R=2$ and extracting the coefficient multiplying $R^2 \pa \theta \bar\pa \theta$ at $R=2$, one finds the effective coupling ${\D R\over 2\pi}$, which plays the role of ${s\over \sqrt{N}}$ in the surface defect problem. It then follows immediately that expanding \eqref{bosans} at small $\D R$ reproduces \eqref{CPTchiral}.

We conclude the conformal perturbation theory analysis around the extraodinary fixed points with two comments.

First, one can in principle carry out a similar analysis for other defect CFT observables, such as the anomalous dimensions of the pseudoscalar and fermion operators that \textit{do not} decouple at the extraordinary fixed points $\mu=\pm {1\over 2}$, as well as the changes in OPE coefficients, all of which depend on $s$ (and hence on $\mu$). In general, however, even the leading corrections at small $s$, to leading order in $1/N$, given by analogs of the integral in \eqref{CPTintegral}, involve nontrivial boundary four-point functions that themselves receive contributions from bulk interactions at the same order (see footnote \cite{myfootnote} for related remarks and \cite{Carmi:2018qzm} for related calculations). By focusing instead on the anomalous dimension of the chiral fermion $\chi^I$, which decouples at the extraordinary point, we avoid having to compute such four-point functions on the normal boundary $|B_1\ra$, since the correlators entering \eqref{CPTintegral} are fixed by free propagators to this order. Nevertheless, it would be very interesting to determine these boundary correlators for $|B_1\ra$, as they encode nontrivial information about the operator data of the normal boundary condition and would provide further consistency checks of the exact solutions obtained here.

Second, and relatedly, the perturbative anomalous dimension of $\chi^I$ in \eqref{CPTchiral}, when compared with the exact large-$N$ result for $\Psi^I_{\rm e}$ in \eqref{fermlargeNanom} using the identification between $s$ and $\mu$ in \eqref{smu}, suggests that, in a suitable scheme, conformal perturbation theory in this problem may be ``two-loop'' exact. Establishing this would require intricate cancellations involving nontrivial boundary correlators of $\Psi^I_{L,R}$.

\end{document}